\documentclass[useAMS, usenatbib]{mn2e}
\usepackage[T1]{fontenc}
\usepackage[utf8]{inputenc}
\usepackage[english]{babel}
\usepackage{rotating}
\usepackage{url}
\usepackage{amsmath}
\usepackage{commath}
\usepackage{amssymb}
\usepackage{graphicx}
\usepackage{subcaption}
\usepackage{pdfpages}
\usepackage{esint}
\usepackage[authoryear]{natbib}
\usepackage{listings}
\usepackage{textcomp}
\usepackage[export]{adjustbox}
\usepackage{booktabs}
\usepackage{siunitx} 
\usepackage{ulem}

\makeatletter

\def\gtsima{$\; \buildrel > \over \sim \;$}
\def\ltsima{$\; \buildrel < \over \sim \;$}
\def\prosima{$\; \buildrel \propto \over \sim \;$}
\def\gsim{\lower.5ex\hbox{\gtsima}}
\def\lsim{\lower.5ex\hbox{\ltsima}}
\def\simgt{\lower.5ex\hbox{\gtsima}}
\def\simlt{\lower.5ex\hbox{\ltsima}}
\def\simpr{\lower.5ex\hbox{\prosima}}
\def\la{\lsim}
\def\ga{\gsim}

\newcommand{\be}{\begin{eqnarray}}
\newcommand{\ee}{\end{eqnarray}}
\def\lsim{\,\lower2truept\hbox{${< \frac\hbox{\raise4truept\hbox{$\sim$}}}$}\,}
\def\gsim{\,\lower2truept\hbox{${> \frac\hbox{\raise4truept\hbox{$\sim$}}}$}\,}

\providecommand{\tabularnewline}{\\}
\title[High-z scaling relations]{The assembly of dusty galaxies at $z \geq 4$: the build-up of stellar mass and its scaling relations with hints from early JWST data}
\author[Di Cesare et al.]{C. Di Cesare$^{1,2,3,4}$\thanks{E-mail:claudia.dicesare@uniroma1.it}, L. Graziani$^{1,2,3}$\thanks{E-mail:
luca.graziani@uniroma1.it}, R. Schneider$^{1,2,3}$, M. Ginolfi$^{4,5}$, A. Venditti$^{1,2,3,6}$,\newauthor P. Santini $^{3}$,  L. K. Hunt$^{7}$  \\
$^{1}$Dipartimento di Fisica, Sapienza, Universit$\grave{a}$ di Roma, Piazzale Aldo Moro 5, 00185, Roma, Italy\\
$^{2}$INFN, Sezione di Roma I, Piazzale Aldo Moro 2, 00185 Roma, Italy\\
$^{3}$INAF/Osservatorio Astronomico di Roma, Via di Frascati 33, 00078 Monte Porzio Catone, Italy\\
$^{4}$European Southern Observatory, Karl-Schwarzschild-Str. 2, D-85748, Garching, Germany \\
$^{5}$Dipartimento di Fisica e Astronomia, Università di Firenze, Via G. Sansone 1, 50019, Sesto Fiorentino (Firenze), Italy \\
$^{6}$Dipartimento di Fisica, Tor Vergata, Universit$\grave{a}$ di Roma, Via Cracovia 50, 00133, Roma, Italy \\
$^{7}$INAF/Osservatorio Astrofisico di Arcetri, Largo E. Femi 5, 50125 Firenze, Italy 
}

\begin{document}

\date{July 2019}

\pagerange{\pageref{firstpage}--\pageref{lastpage}} \pubyear{2016}

\maketitle

\label{firstpage}

\begin{abstract}
The increasing number of distant galaxies observed with ALMA by the ALPINE and REBELS surveys and the early release observations of the JWST promise to revolutionize our understanding of cosmic star formation and the assembly of normal, dusty galaxies. Here we introduce a new suite of cosmological simulations performed with \texttt{dustyGadget} to interpret high-redshift data. We investigate the comoving star formation history, the stellar mass density and a number of galaxy scaling relations such as the galaxy main sequence, the stellar-to-halo mass and dust-to-stellar mass relations at $z > 4$. The predicted star formation rate and total stellar mass density rapidly increase in time with a remarkable agreement with available observations, including recent JWST ERO and DD-ERS data at $z \geq 8$. A well defined galaxy main sequence is found already at $z < 10$ following a non evolving power-law, which - if extrapolated at high-mass end - is in agreement with JWST, REBELS, and ALPINE data. This is consistent with a star formation efficiently sustained by gas accretion and a specific star formation rate increasing with redshift, as established by recent observations. A population of low-mass galaxies ($8 < \rm{Log(M_\star/M_\odot)} < 9$) at $z \leq 6 - 7$ that exceeds some of the current estimates of the stellar mass function is also at the origin of the scatter in the stellar-to-halo mass relation. Future JWST observations will provide invaluable constraints on these low-mass galaxies, helping to shed light on their role in cosmic evolution.

\end{abstract}


\section{Introduction}

Since the Atacama Large Millimeter Array (ALMA)\footnote{\url{http://www.almaobservatory.org}} started observing the Universe at the highest redshifts, our view of the first stages of cosmic star formation and galaxy assembly has significantly improved, and we discovered that within the first 1.5 billion years of galaxy evolution ($z>4$) the process of cosmic star formation had a profound and immediate impact on the chemical evolution of young galaxies \citep{BrommYoshida2011}. 

Although deep observations of single, high-redshift dusty galaxies at the Epoch of Reionization (EoR) significantly increased in the last decade \citep{Cooray+2014,Watson+2015,Capak+2015,Hashimoto+2018,Laporte+2017,Tamura+2019,Bakx2020}, the recent advent of high-redshift surveys such as the ALMA Large Program to Investigate [CII] at Early Times Survey (ALPINE\footnote{\url{http://alpine.ipac.caltech.edu}},  \citealt{LeFevre+2020,Faisst+2020,Bethermin+2020}) and the Reionization Era Bright Emission Line Survey (REBELS, \citealt{Bouwens2022}) has opened up the possibility to build a coherent view of the early stages of galaxy evolution and to explore the early evolution of galaxy scaling relations, observationally well established at lower redshifts (see for example \citealt{Zahid2013,Cresci2019,Ginolfi2020,Hunt2020,Casasola2020,Kumari2021,Tortora2022,Hayden-Pawson2022}). The combined dataset of the above surveys covers in fact two complementary redshift ranges: $4.4 \leq z \leq 5.9$ (ALPINE) and $6.5 \leq z \leq 9.4$ (REBELS), and already revealed the presence of chemically evolved, highly interacting galaxies in the early universe, with hints on an unexpected population of dusty, obscured star-forming objects \citep{Fudamoto+2021}. Even more intriguing, a recent analysis of a limited sample of galaxies available at even higher redshifts ($z \geq 9$, \citealt{Tacchella+2022}) provides indications of efficient metal production at the early stages of cosmic reionization. 

The exciting, early release observations of the JWST have already provided evidence of a significant star formation activity at $z > 11$ \citep{Adams+2022,Atek+2022,Castellano2022,Donnan+2022,Harikane+2022,Naidu2022,Yan+2022,Zavala+2022}, with candidate galaxies showing a variety of physical properties \citep{Leethochawalit2022, Santini2022}. Another interesting candidate is found at $z \sim 14$ with $\rm{Log(M_\star/M_\odot)} \sim 8.5$ \citep{Filkenstein2022}, in addition to 7 massive objects with $\rm{Log(M_\star/M_\odot)} > 10$ at $7 < z < 11$, including two galaxies with a surprisingly high stellar mass of  $\rm{Log(M_\star/M_\odot)} \sim 11$ at these early epochs \citep{Labbe2022}, then questioned by \citet{Steinhardt+2022}.  Although preliminary and still not spectroscopically confirmed, these early results suggest an early onset of galaxy evolution, consistent with the picture outlined at longer wavelengths by the ALPINE and REBELS surveys.

The ALPINE collaboration provided the first comprehensive, statistically significant, multi-wavelength (from rest-frame UV to the far-infrared) sample of 118 spectroscopically selected main sequence galaxies evolving at the end of the Epoch of Reionization. More information  on the detection strategy, the data-processing and the ancillary data can be found in \citet{LeFevre+2020,Bethermin+2020, Faisst+2020}. The ALPINE sample targets the emission of single ionised carbon [CII] at $158 \mu$m, which traces both emission from star-forming regions and molecular hydrogen gas-clouds; the thermal continuum from dust emission is also available for a wide set of galaxies observed in the redshift range $4.4 < z < 5.9$. Using the ALPINE sample, both scaling relations and single objects properties have been deeply investigated: the star formation rate density (computed from the UV+IR emission), the main sequence and the specific star formation rate relations are discussed in \citet{Khusanova+2021}, while \citet{Pozzi+2021} investigated the dust-to-stellar mass relation; the star formation rate density from the total IR luminosity function is finally estimated by  \citet{Gruppioni+2020}. A careful analysis of the kinematic diversity and rotation of massive star-forming objects can be found in \citet{Jones+2021}; \citet{Ginolfi+2020b} focused on the pollution of the circumgalactic medium of a merging system, while an interesting case of a triple merger at $z \sim 4.56$ is discussed in \citet{Jones+2020}. The aim here is just to mention few among the many works published by the ALPINE collaboration, and should not be  considered as an exhaustive and complete list. Finally, ALPINE observations first revealed that a significant fraction of star formation in the post-reionization epoch is already hidden by dust clouds \citep{ESOsummary+2020}. 

The REBELS survey \citep{Bouwens2022} complements the ALPINE sample by dramatically increasing the number of spectroscopically confirmed galaxies and dust-continuum detections at $z > 6.5$. REBELS targets a photometrically selected sample of 40 UV-bright galaxies from a number of fields including COSMOS/UltraVISTA, VIDEO/XMM-LSS+UKIDSS/UDS, HST legacy fields, and the BoRG/HIPPIES pure parallel fields. More details on the observational strategy and statistical significance of the REBELS detections can be found in \citet{Bouwens2022} (but see also \citealt{Smit2018, Schouws2022b, Schouws2022a}). REBELS observations of the above candidates already tripled the number of ISM cooling lines ([CII]158$\mu$m, [OIII]88$\mu$m, Schouws et al. in prep) and dust continuum detections of galaxies found in the Epoch of Reionization \citep{Inami2022}, allowing us to explore the nature of dust-rich galaxies, to characterise their dust properties, and to study the dust buildup at these early cosmic epochs (e.g. \citealt{Dayal2022, Ferrara2022, Sommovigo2022}, Schneider et al., in prep., Graziani et al., in prep.). Detections of a strong Ly$\alpha$ line associated with the largest [CII] line widths present in some candidates at $z \sim 7$ is discussed in \citealt{Endsley2022}. Finally, for an extended discussion on the specific star formation rates (sSFR) of all the galaxies in the sample, the interested reader is referred to \citet{Topping+2022}.

Interestingly, both surveys probe objects with clear detections of common lines and dust continua as well as consistent absolute UV magnitudes (from -21.3 to -23 in REBELS and -20.2 to -22.7 in ALPINE). Compatible ranges of stellar masses are also found, allowing us to trace galaxy properties across the redshift range $4 \leq z \leq 7$ and to extensively compare with galaxy formation models.

During the past few years, many studies which were based on data constrained models \citep{Imara+2018,Behroozi_2019}, classical semi-analytic methods  \citep{Popping+2017,Somerville_2018,Yates2021,Trinca+2022}, semi-numerical models running on halo merger histories extracted from N-body simulations \citep{Mancini2015,GrazianiGAMESH2017,GinolfiGAMESH2018,UcciAstraeus2021, WangUMachine2021}, or hydrodynamical simulations \citep{ Sarmento_2018, Pallottini2019, Pallottini2022, Graziani+2020, KannanThesan2022, Wilkins+2022} investigated the high-redshift Universe. This has been done with the purpose of either interpret datasets based on limited observational samples or to provide forecasts for JWST observations \citep{Yung+2019,BehrooziJWST2020}. The availability of coherent observational samples from surveys certainly offers remarkable advantages to cosmological models as they allow (i) to constrain the properties of a wide range of simulated galaxies discovered in models at progressively high-mass resolution; (ii) to assess the impact of feedback processes on galaxy evolution, (iii) to discover a possible redshift evolution in scaling relations well known at lower redshift. Although at low redshift ($z\leq5$) the evolution of stellar mass functions is relatively consolidated \citep{Baldry+2012, Tomczak+2016, Adams+2021, Leslie+2020}, at higher redshifts disagreements emerge \citep{Oesch+2014,Bhatawdekar+2019, Kikuchihara+2020,Stefanon+2021, Harikane+2022}. In addition, dust obscuration complicates the measurement of the SFR, leading to an uncertain picture of the cosmic star formation rate density (CSFRD) above $z\geq2$ \citep{Casey2018, Gruppioni+2020, Zavala2021, Barrufet+2022}. Such discrepancies suggest that at early times, the physical processes that regulate galaxy evolution are still not completely understood. Simulations and their comparison with observations provide an effective path forward, in order to better constrain galaxy properties and their evolution in the early Universe.

In this paper we introduce a new suite of eight statistically independent hydrodynamical simulations evolving cosmic volumes of 50$\rm{h}^{-1}$~cMpc / side length with a common chemical and mechanical feedback model; all runs are performed with the \texttt{dustyGadget} code \citep{Graziani+2020} improving the mass resolution and statistical significance of the original work. The new simulation suite provides then a large sample of dusty galaxies suitable to: (i) investigate fundamental galaxy scaling relations (e.g the galaxy main sequence, the stellar mass function, the specific star formation rate evolution, the dark matter halo mass-to-stellar mass etc...) at $z \geq 4$; (ii) have access to a rich set of dusty halo environments in which different galaxy populations assemble and evolve (Schneider et al., in prep; Graziani et al., in prep); (iii) explore the nature of the stellar populations and star-forming regions hosted by the brightest systems at $z > 6$ (Venditti et al., in prep). 

In this paper, we use the new \texttt{dustyGadget} simulation suite to investigate galaxy scaling relations at $z \geq 4$ as probed by current data at different wavelengths, including the early JWST observations. The redshift evolution of all the above objects and their placement on scaling relations allow to firmly connect the core data of the REBELS and ALPINE samples in a coherent evolutionary model and will serve to disentangle odd and more rare galaxy evolution histories (Di Cesare et al., in prep.).

The paper is organized as follows: in Section \ref{sec:galForm} we introduce the \texttt{dustyGadget} model and the new simulation suite, while Section \ref{sec:results} discusses the results of our analysis. In Section \ref{sec:SFRDeSMD} we discuss predictions of the cosmic star formation rate density and the stellar mass density, in Section \ref{sec:StellarMF} we explore the stellar mass functions. The canonical galaxy scaling relations are also investigated: the galaxy main sequence is discussed in Section \ref{sec:MS} and \ref{sec:AddMS}, the specific star formation rate evolution in Section \ref{sec:sSFR}, the relation between dark matter halo mass and stellar mass in \ref{sec:halo-stellar} and the $\rm{M_{dust}-M_{\star}}$ relation in Section \ref{sec:dust-star}. Finally, Section \ref{sec:Conclusions} draws our conclusions.

\section{Galaxy formation simulations}
\label{sec:galForm}
This section describes the hydrodynamical code \texttt{dustyGadget} \citep{Graziani+2020}, which extends the original implementation of  \texttt{Gadget} \citep{Springel2005} and its successive improvements  \citep{Tornatore+2007a, Tornatore+2007b, Maio+2009} by implementing a model of dust production and evolution in the interstellar medium (ISM) of the simulated galaxies, consistent with the two-phase model of \citet{SpringelHernquist2003}. The code also follows the spreading of grains and atomic metals through galactic winds at the scales of both circumgalactic and intergalactic medium (CGM/IGM). 

The chemical evolution model of \texttt{dustyGadget} for the gaseous components derives from the original implementation of \citet{Tornatore+2007b}: the model relaxes the \textit{Instantaneous Recycling Approximation} (IRA) and follows the metal release from stars of different masses, metallicity and lifetimes. Different mass and metallicity-dependent yields are implemented for PopII/I stars: coming either from core-collapse supernovae (SNe) or type-Ia supernovae (SNIa). Stars with masses $\ge 40 \: \rm{M_{\odot}}$ are assumed to collapse into black holes and do not contribute to metal enrichment. PopIII stars with masses $140 \: \rm{M_{\odot}} \le \rm M_{\star} \le 260 \: \rm{ M_{\odot}}$ are expected to explode as pair-instability SNe (PISN), according to mass dependent yields from \citet{Heger+2002}. PopIII stars which masses lie outside the PISN mass range are assumed to collapse into black holes. The chemical network in \texttt{dustyGadget} also includes the evolution of both atomic and ionized hydrogen, helium, and deuterium by relying on the standard \texttt{Gadget} implementation of the cosmic UV background, first introduced in \citet{Madau1996}. The interested reader can find more details in \citet{Graziani+2020} and references therein.

Cosmic dust is introduced in the previous chemical network consistently with the ISM cold and hot phases. Dust production by stars is implemented to ensure consistency with gas phase metal enrichment: mass and metallicity-dependent dust yields \citep{BianchiSchneider2007, Marassi2019} are computed for different stellar populations (PopII, PopI and core-collapse SNe, PopIII and PISN) and eventually corrected for the effects of the reverse shock process occurring at unresolved scales \citep{Bocchio2014,Bocchio2016}. Following \citet{Graziani+2020}, four grain species are modelled: Carbon (C), Silicates $\rm{(MgSiO_3,Mg_2SiO_4,SiO_2)}$, Aluminia $\rm{(Al_2O_3)}$ and Iron (Fe) dust. However, the chemical evolution model is flexible enough to include other grain types and to explore combinations of stellar yields and different assumptions on the shape of the stellar Initial Mass Function (IMF). Once the grains produced by stars are released into the ISM, according to the properties of the environment in which they evolve, they experience different physical processes altering their mass, relative abundances, chemical properties, charge, and temperature. It is generally assumed that the dust-to-light interactions (e.g. photo-heating, grain charging) change the thermodynamic and electrical properties of the grains (see for example \citealt{Glatzle2019,Glatzle2022}) but have a negligible impact on the total dust mass unless the grain temperatures reach the sublimation threshold ($\rm{T_{d,s}} \gtrsim 10^{3}$~K). Other physical processes (i.e. sputtering and grain growth\footnote{Note that a subtle interplay between grain charging and grain growth process could alter the efficiency of the latter, as discussed in \citet{Glatzle2022}.}) can alter, on the other hand, both the total dust mass and the grain size distribution \citep{Draine+2011,Aoyama2020}. The last version of \texttt{dustyGadget} does not take into account the evolution of grain sizes, but it only considers physical processes which directly alter the dust mass e.g. grain growth, destruction by interstellar shocks, and grain sputtering in the hot ISM phase (see \citealt{Graziani+2020} for more technical details on their numerical implementation).

Finally, at the end of stellar evolution metals and dust are spread in the surroundings of star-forming regions. The dust distribution follows the atomic metal spreading without accounting for any momentum transfer through dust grains. At the same time, dusty particles associated with galactic winds evolve in their hot phase through sputtering. Therefore, the dust-to-metal ratio will be modulated depending on the environment, obtaining different values for the galactic ISM, CGM and IGM.

A new suite of eight statistically independent cosmological simulations, hereafter dubbed as U6-U13, is adopted in the present work. All the runs share a common physical setup and simulate cosmic volumes of $50\rm{h}^{-1}$~cMpc side length adopting a flat $\Lambda$CDM cosmology with $\Omega _{\Lambda} = 0.6911$, $\Omega _m = 0.3089$, $\Omega _b = 0.0486$ and $h=0.6774$, consistent with Planck 2015 (\citealt{PLANK+2016}).
The eight cosmological simulations have an equal mass resolution of $3.53 \times 10^7 \rm{h}^{-1} \: \rm{M_\odot}$ for dark matter (DM) particles and of $5.56 \times 10^6 \rm{h}^{-1}\: \rm{M_\odot}$ for gas particles, setup with $2 \times 672^3$ total number of particles\footnote{Note that both volume and the mass resolution of the present simulation is increased with respect to the one discussed in \citet{Graziani+2020}.}. The cubic volume and SPH resolution are chosen to guarantee a good compromise between an adequate statistics in each run and an acceptable number of galaxies resolved with a total stellar mass $\rm{Log(M_{\star}/M_{\odot})} \geq 8$ by $\rm{N_{\star}} > 54$ stellar particles of individual mass $1.3 \times 10^6 \rm{h}^{-1} \: \rm{M_\odot}$. In addition, all the above requirements ensure a reasonable computational time for eight runs performed in the redshift range $4 \leq z \leq 100$. A good statistics of galaxy candidates is required in fact to reproduce a reliable trend of cosmic star formation history and the main galaxy scaling relations, while the ability to resolve the most massive star-forming environments is fundamental to perform a first exploration of the internal properties of some of these candidates and their circumgalactic environments and compare with possible observational counterparts\footnote{Once the best candidates are chosen we plan to perform zoom-in simulations of their environments with the next version of \texttt{dustyGadget} providing an updated model for H$_{2}$-based star formation and ISM (Graziani et al., in prep.).}.

To ensure consistency across the simulations, all the volumes share common assumptions on star formation prescriptions, mechanical, chemical and radiative feedback processes as described in \citet{Graziani+2020}; in the following paragraph we briefly recap the main physical setup of the simulation. 
The simulations start at $z=100$ assuming neutral pristine gas and evolve all particle components down to $z = 4$ with 40 outputs at intermediate redshifts. For a better comparison with previous work, and to assess the statistical convergence of the previously studied relations, we adopted a chemical network accounting for molecules and atomic metals (see \citealt{Maio+2010} for more details). Star formation occurs in the cold phase of gas particles once their density exceeds a value of $n_{\rm th} = 132 \, \rm{h}^{-2} \rm cm^{-3}$ (physical)\footnote{This choice allows to capture all the relevant phases of cooling until the onset of runaway collapse, as discussed in \citet{2009A&A...503...25M}.}. The IMF of the stellar populations, each represented by a single stellar particle, is assigned according to their metallicity $\rm{Z_\star}$, given a gas critical metallicity $\rm{Z_{crit}} = 10^{-4} \: \rm{Z_\odot}$\footnote{Here, we assume $\rm{Z_\odot} = 0.02$ \citep{Grevesse&Anders1989}.}. When $\rm{Z_\star < Z_{crit}}$ we adopt a Salpeter IMF \citep{Salpeter+1955} in the mass range $[100 - 500]\: \rm{M_\odot}$. Otherwise, the stars are assumed to form according to a Salpeter IMF in the mass range $[0.1 - 100] \: \rm{M_\odot}$. Galactic winds are modelled with a constant velocity of $500~\textrm{km}\,\textrm{s}^{-1}$, in line with outflows observed in ALPINE normal galaxies \citep{2020A&A...633A..90G}. Radiative feedback is implemented instead as in the original version of \texttt{Gadget}, i.e. by adopting a cosmic UV background \citep{1996ApJ...461...20H}. Apart from the aforementioned calibrations (i.e. galactic winds, radiative feedback), our model is not calibrated on any particular observational set or survey. We warn the reader that our simulations do not model the formation of Active Galactic Nuclei (AGN)\footnote{The AGN feedback is presumed to be responsible for the bending at the high mass end of the main sequence and, since we do not model the formation of AGNs, we find a different shape for such relation if compared to models in which the AGN feedback is taken into account (see Appendix \ref{sec:AppMS}).} and do not account for mechanical or radiative feedback of formed black holes.

Finally, a common post-processing setups is adopted as well, in order to identify DM halos and their substructures trough the \texttt{AMIGA} halo finder (AHF, \citealt{Knollmann+2009}). 

\section{Results}
\label{sec:results}
Here we discuss the results of our simulations. Section \ref{sec:SFRDeSMD} investigates cosmic star formation and the galaxy stellar mass function is derived in Section \ref{sec:StellarMF}. Then we introduce many galaxy scaling relations connecting the stellar mass of collapsed objects (M$_{\star}$) with other galaxy or DM halo properties, including the main sequence of galaxy formation in Section \ref{sec:MS} and the specific star formation rate (sSFR-z) in Section \ref{sec:sSFR}. Finally, the relation connecting stellar and DM halo mass is discussed in Section \ref{sec:halo-stellar}, while the connection with the dust mass (M$_{\rm d}-$M$_{\star}$) is investigated in Section \ref{sec:dust-star}.

\subsection{Cosmic star formation history and cosmic stellar mass density} 
\label{sec:SFRDeSMD}

The history of cosmic star formation, i.e. the redshift evolution of the total CSFRD ($\Psi$) and/or the total stellar mass density (SMD, $\rho_\star$) are discussed in this section, comparing the predictions of \texttt{dustyGadget} with available observations and recent theoretical models at $z \geq 4$. The relations investigated here account for quantities directly inferred from gas and stellar particles integrated into a comoving volume V$_\mathrm{C} = (50\rm{h}^{-1}$)$^3$~cMpc$^3$ without requiring any halo/galaxy definitions. 

\begin{figure*}
\centering
\includegraphics[angle=0,width=0.9\textwidth,clip]{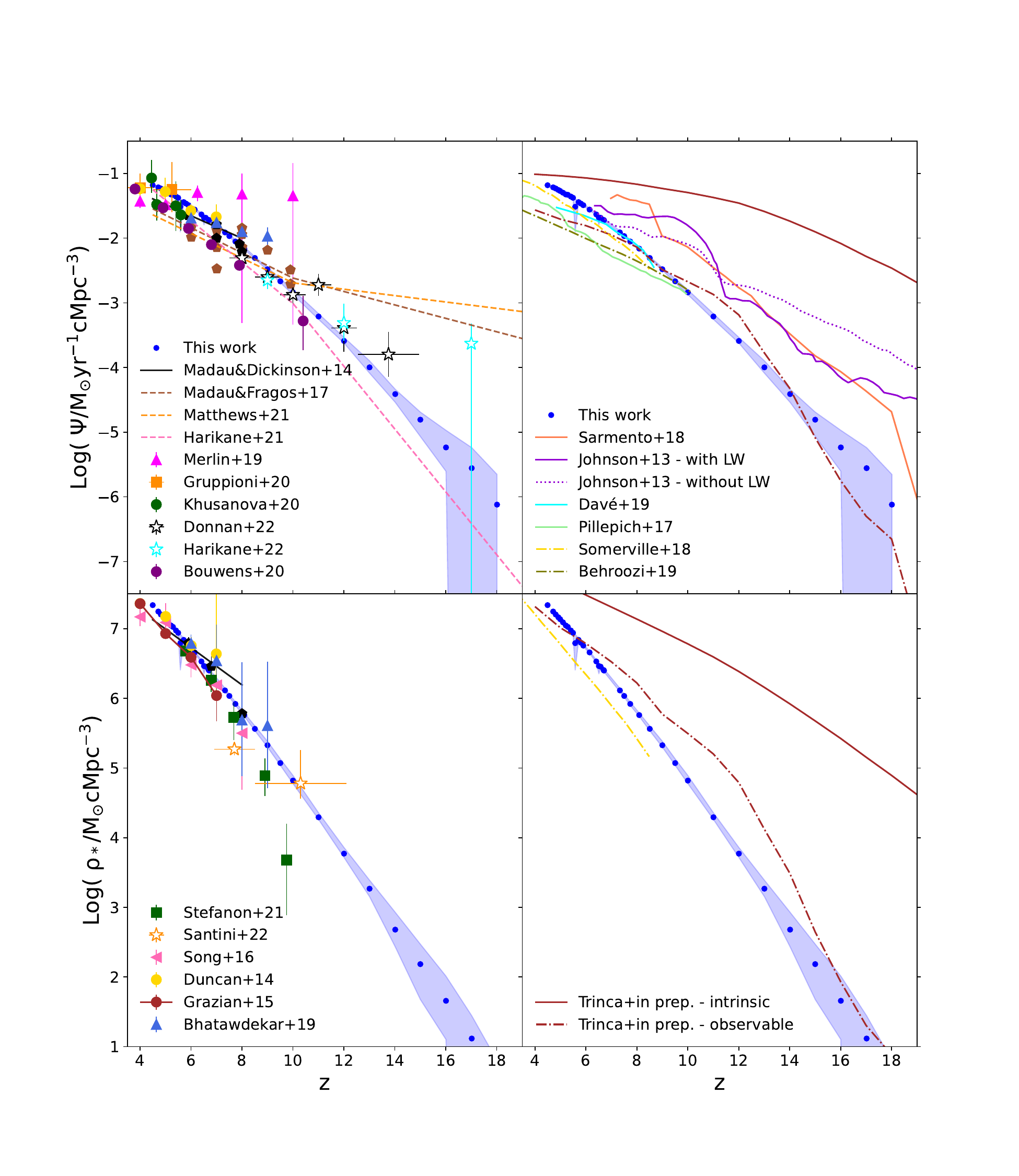}
 \caption{Cosmic SFRD $\Psi$ (\textbf{top panels}) and SMD $\rho_*$ (\textbf{bottom panels}) as a function of redshift $z$ in the range $4 \leq z \leq 19$ and averaged through U6-U13 (see text for more details). \textbf{Left panels:} mean value of $\Psi$ and $\rho_*$ (\textit{blue dots}) and their spread between min/max found in the eight cubes (\textit{blue shaded areas}). Observed points are taken from \citet{Madau&Dickinson_2014} (\textit{black pentagons}), \citet{Madau&Fragos_2017} (\textit{sienna pentagons}), \citet{Bouwens_2020} (\textit{purple dots}), \citet{Gruppioni+2020} (\textit{orange squares}), \citet{Khusanova+2021} (\textit{green dots}), \citet{Merlin+2019} (\textit{magenta triangles}), \citet{Stefanon+2021} (\textit{green squares}), \citet{Duncan+2014} (\textit{yellow dots}), \citet{Bhatawdekar+2019} (\textit{sky-blue triangles}), \citet{Song+2016} (\textit{pink triangles}), \citet{Donnan+2022} (\textit{empty stars with black border}), \citet{Harikane+2022} (\textit{empty stars with cyan borders}) and \citet{Santini2022} (\textit{empty stars with orange border}). The data constrained best-fit functions from \citet{Madau&Dickinson_2014} are shown as \textit{black solid lines}, while the fits from \citet{Madau&Fragos_2017}, \citet{Matthews_2021} and \citet{Harikane_2021} are shown as \textit{sienna/dark orange/pink dashed lines}; notice that from $z=10$ these are extrapolations. \textbf{Right panels}: Comparison with the results from other models and simulations, when available: \citet{Sarmento_2018} (numeric simulation, \textit{coral solid line}), \citet{Johnson_2012} (\texttt{FiBY} simulation, \textit{violet lines} - the simulations with and without LW flux are shown respectively in \textit{solid} and \textit{dotted} line styles), \citet{Dave_2019} (\texttt{SIMBA} simulation, \textit{cyan solid line}), \citet{Pillepich_2017} (\texttt{IllustrisTNG} simulation, \textit{light green solid line}), Trinca et al., in prep (\texttt{CAT} semianalytic model \textit{brown lines} - intrinsic and observable models are respectively in \textit{solid} and \textit{dashed-dotted} line styles), \citet{Somerville_2018}, (semianalytic model, \textit{golden dashed dotted line}), \citet{Behroozi_2019} (\texttt{UNIVERSEMACHINE} semianalytic model, \textit{olive dash dotted line}).}
\label{fig:SFRD_SMD}
\end{figure*}

For each cube at a given redshift $z$, $\Psi(z)$ is computed as:

\begin{equation}
    \Psi (z) = \frac{\sum_i \mathrm{SFR}_i(z)}{V_\mathrm{C}},
    \label{eq:SFRD}
\end{equation}
\noindent
where SFR$_i$ is inferred from the \textit{i}-th star-forming gas particle in the cube. While $\rho_*(z)$ is defined as:

\begin{equation}
    \rho_\star (z) = \frac{\sum_i M_{\star,i}(z)}{V_\mathrm{C}},
    \label{eq:SMD}
\end{equation}
\noindent
where $\rm{M_{\star,i}}$ is the total mass of the \textit{i}-th stellar particle in the cube. Both quantities are shown in Figure \ref{fig:SFRD_SMD} in the redshift range $4 \leq z \leq 19$. Blue, filled dots indicate the mean values at any given redshift among the volumes (U6 - U13), while the shaded areas show the minimum-maximum spread found across the whole simulation sample.  $\Psi(z)$ rapidly increases with decreasing $z$, from $\Psi \sim 10^{-6} - 10^{-7}$~M$_\odot$ yr$^{-1}$ cMpc$^{-3}$, up to $ \Psi\sim 10^{-2} - 10^{-1}$~M$_\odot$ yr$^{-1}$ cMpc$^{-3}$ by $z \sim 4$; this trend is mirrored by $\rm{\rho_\star (z)}$ in all the volumes, as the stellar mass accumulates across cosmic time. In the same redshift range in fact, $\rm{\rho_\star (z)}$ increases by more than 6 orders of magnitude, starting from $\rm{\rho_\star \sim 10 \; \si{M_\odot.cMpc^{-3}}}$. 

To better understand the scatter across the eight cubes, we investigate the values relative to each individual simulation finding a tight convergence starting at $z \sim 12$. Their spread becomes relevant instead at $z \gtrsim 12 $, certainly because of the cosmic variance. At $z \gtrsim 12$, the number of star-forming systems is too scarce to collect detailed statistics from a single simulated volume, therefore their quantities hardly reflect the cosmological mean value. For this reason, having more than one cube is effective in increasing the global statistics at Cosmic Dawn. Despite the above improvement, star formation at these early times is not yet robustly structured and its evolution remains strongly dependent on both different initial conditions and assembly histories of each cube. In addition, our mass resolution does not allow us to resolve the first star forming regions in minihalos, and the CSFRD and SMD are likely to be underestimated at the highest redshifts.
 
To verify the above predictions, in Figure \ref{fig:SFRD_SMD} we compare them with the expectations of the data-constrained model of \citet{Madau&Dickinson_2014} and \citet{Madau&Fragos_2017}, and with other estimates of the SFRD based on IR \citep{Khusanova+2021,Gruppioni+2020,Matthews_2021,Merlin+2019}, rest frame UV \citep{Donnan+2022} and UV+dust corrected \citep{Bhatawdekar+2019,Bouwens_2020,Duncan+2014,Harikane_2021,Harikane+2022} observations (top left panel). The very recent JWST results are those from \citet{Donnan+2022} and \citet{Harikane+2022}. The SMD is compared instead with \citet{Madau&Dickinson_2014,Stefanon+2021,Song+2016,Duncan+2014, Grazian2015,Bhatawdekar+2019} and the latest JWST estimates by \citet{Santini2022} (bottom left panel). Our results for both SFRD and SMD relations are in overall agreement with \citet{Madau&Dickinson_2014} at $z \lesssim 8$, even though our trends appear slightly steeper at decreasing redshifts, but still in excellent agreement with \citet{Duncan+2014} and recent ALPINE estimates in \citet{Gruppioni+2020} and \citet{Khusanova+2021}. 

At higher redshifts ($z \ge 8$) the values predicted by \texttt{dustyGadget} are remarkably consistent with the recent results from JWST Early Release Observations (ERO) and Early Release Science Program (ERS) both for the SFRD \citep{Donnan+2022} and the preliminary estimates of the observed SMD \citep{Santini2022}. This points to a higher rate of star formation at high-redshift than previously indicated by the ALMA Spectroscopic Survey Large Program \citep{Bouwens_2020}, \citet{Harikane_2021} (top left panel, in purple dots and pink dashed lines) and by \citet{Stefanon+2021} (bottom panel, green squares). 

In the right panels of Figure \ref{fig:SFRD_SMD}, we compare our results with some predictions of other semi-analytic models and simulations i.e. \citet{Sarmento_2018,Johnson_2012,Dave_2019, Pillepich_2017,Somerville_2018, Behroozi_2019}, Trinca et al. (in prep). In particular, for the Cosmic Archaeology Tool (\texttt{CAT}, described in \citealt{Trinca+2022}) we show the estimates for the intrinsic and observable SFRD and SMD: this last estimate has been obtained using a threshold at $\rm{M_{UV}} < -17.5$. These models predict quite different trends, especially at high-redshifts ($z>8$), indicating that star formation at very early times is strongly dependent on the analysed cosmological volume and/or the adopted feedback model.

\subsection{The stellar mass function}
\label{sec:StellarMF}

The growth of stellar mass (M$_{\star}$) during galaxy assembly is often investigated with the stellar mass function ($\Phi$, SMF) as it provides important hints on how the total stellar mass present in a cosmic volume (see the previous section) distributes across different luminous structures. $\Phi$ is usually defined as the number density of galaxies as a function of their M$_{\star}$, collected in a fixed redshift interval. 

Here we investigate the evolution of $\Phi(z)$ in the redshift range $z=10-4$ as predicted by \texttt{dustyGadget} runs computed as:

\begin{equation}
\Phi [{\rm dex^{-1} Mpc^{-3}}]=\frac{d \rm{N_i}}{d{\rm Log}\rm{M_{\star}}}\frac{1}{\rm V_{\rm C}},
\end{equation}
\noindent
where $N_i$ is the number of galaxies in the $i$-th $\rm{Log(M_{\star}/M_\odot)}$ bin and V$_{\rm C} = 50^3$h$^{-3}$~cMpc$^3$ is the simulated comoving volume. In each simulated universe, galaxies are extracted from available catalogs at common, fixed redshifts and are selected in the mass interval $\rm{Log(M_{\star}/M_{\odot})} \in [8 - 11]$\footnote{The lower bound of the mass interval is set to properly resolve low-mass objects and the upper bound is linked to the size of the simulated volume.}, then binned within 0.2 dex. The resulting value in each bin is finally divided by the size of the bin and V$_{\rm C}$. 

\begin{figure*}
\centering
\includegraphics[width=\textwidth]{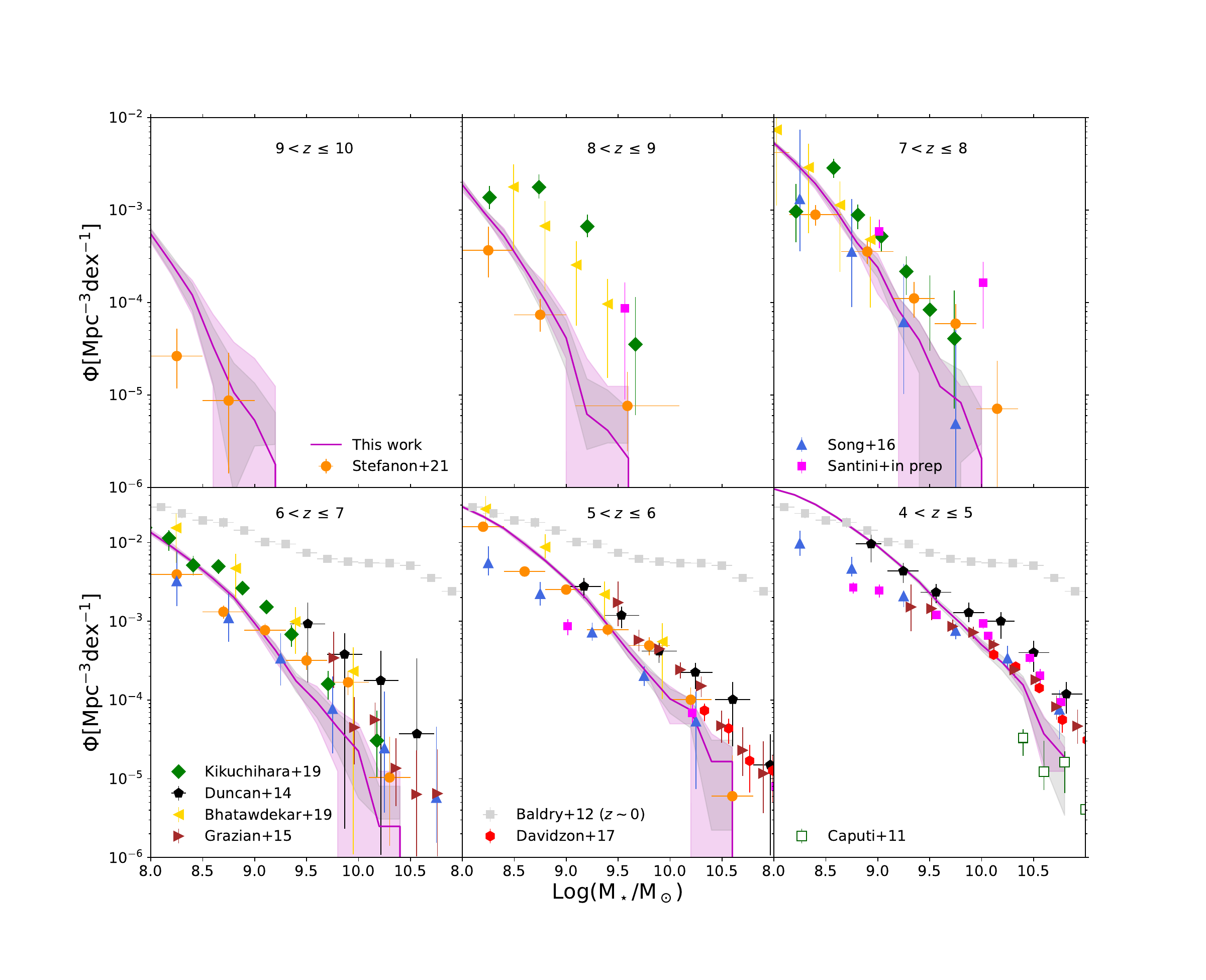}
 \caption{Mass Function obtained from our simulated galaxies in the range $z=10-4$: \textit{magenta solid lines} show mean values of our sample, \textit{pink shaded areas} the min-max spread of the simulations and \textit{gray shaded areas} the Poissonian error associated with our sample. The SMF from observed galaxies are also reported in each panel:  \citet{Stefanon+2021,Song+2016,Duncan+2014,Bhatawdekar+2019,Kikuchihara+2020, Grazian2015, Davidzon+17,Caputi+2011} and Santini et al. (in prep) are respectively in \textit{orange circles, blue triangles, black pentagons, yellow triangles, green diamonds, brown triangles, red hexagons, green empty squares and magenta squares}. Finally, observational constraints at $z \sim 0$ (\textit{gray squares}) from \citet{Baldry+2012} are shown at $z<6$ to guide the eye.}

\label{fig:FigSMF}
\end{figure*}

From top left to bottom right, the panels of Figure \ref{fig:FigSMF} compare $\Phi$ in different redshift intervals\footnote{Each panel assumes the mid point of each specified interval as the reference redshift at which the analysis has been conducted.} with available observations. The mean value of $\Phi$ in each mass bin is shown as magenta solid lines, the min-max spread found across the simulated sample  as pink shaded areas, while the Poissonian errors  are shown as gray shaded areas. Furthermore, we apply the proper conversion factors \citep{Madau&Dickinson_2014} to stellar masses that were originally computed with an IMF different from the \citet{Salpeter+1955} one.

In recent years, predictions of the high-redshift SMF both from observed data samples and theoretical models have been published, complementing the estimates available in the Local Universe (see for example the GAMA survey, \citealt{Baldry+2012}).  \citet{Grazian2015} reconstructed the galaxy stellar mass function in the redshift range $3.5 \leq z \leq 7.5$ by collecting data from the CANDELS/UDS, GOODS-South, and HUDF fields also providing a careful analysis of the many sources of uncertainty when deriving $\Phi$: stellar masses of observed galaxies, assumptions on the star formation histories and on the evolution of their metallicity. Random errors and discrepancies originating from the adopted statistical methods and their assumptions are also carefully discussed, such as the impact of nebular lines and the modelled continuum, the effects of cosmic variance and the possible contamination by AGN sources. The predicted SMF is represented in the figure panels as brown triangles and shows a good agreement with the simulated $\Phi$ at $\rm{Log(M_{\star}/M_{\odot})} > 9.5$, while providing lower values for smaller objects observed at $4 < z \leq 5$ (see bottom right panel). Recently Santini et al. (in prep.) extended this analysis by combining all CANDELS fields and the parallel fields from the Hubble Frontier Fields program \citep{Lotz+2017}, whose depth is crucial to probe the highest redshift galaxies. Details on the adopted technique can be found in \citet{Santini+2021} and Santini et al. (2022, subm.). The results, shown in the figure as magenta squares, confirm the previous considerations for $\rm{Log(M_{\star}/M_{\odot})} > 9.5$. To extend the comparison at higher redshifts and provide a more precise indication on the scatter among different observations, we complement the above dataset with estimates from \citet{Duncan+2014} (black pentagons\footnote{These data are based on deep near-infrared observations that were available in the CANDELS GOODS South field.}), \citet{Song+2016} (sky blue triangles), \citet{Bhatawdekar+2019} (yellow triangles), \citet{Kikuchihara+2020} (green diamonds), and \citet{Stefanon+2021} (orange dots). Finally, and as a reference, we show observational constraints obtained in the Local Universe (gray squares)  by analysing the results of the Galaxy And Mass Assembly (GAMA) survey \citep{Baldry+2012}.

As shown in Figure \ref{fig:FigSMF} an overall general agreement between the simulation and the observed samples can be found in the explored redshift range. At the highest redshifts, ($9<z\le 10$, top left panel) the simulation predicts a number density of faint objects that exceeds recent estimates provided by \citet{Stefanon+2021}, but note that the statistical sample of simulated and observed systems with $\rm{Log(M_{\star}/M_{\odot})} > 8$ in this redshift range is less significant than at lower redshift and more data is necessary to consolidate the trends. While at $8<z\leq9$ (top middle  panel) \texttt{dustyGadget} seems to predict a lower number of objects compared with observations of \citet{Kikuchihara+2020,Bhatawdekar+2019} and Santini et al. (in prep.), it agrees with the estimates by \citet{Stefanon+2021}. In the redshift range $4<z\leq 7$ the low-mass end ($\rm{Log(M_{\star}/M_{\odot})} < 9.5$) appears to be overestimated when compared with data in  \citet{Song+2016} and Santini et al. (in prep.), while the agreement improves when all the other observational estimates are accounted for. An opposite trend is found instead in the high-mass end ($\rm{Log(M_{\star}/M_{\odot})} > 9.5$) where the number of objects predicted by our simulations is lower than estimates from observations. This can be due to the adopted volume in \texttt{dustyGadget} simulations which limits the number of high-mass ($\rm{Log(M_\star/M_\odot)} > 10.5$) galaxies. Finally, the interested reader is referred to Figure \ref{fig:FigSMFmod} in Appendix \ref{sec:AppSMF} for a comparison between our predictions and other available theoretical models.

Constraints from JWST observations on the population of luminous galaxies having $\rm{Log(M_{\star}/M_{\odot})} < 9$ will be crucial for theoretical models in order to assess the properties of their metal/dust enriched ISM through emission lines, as well as their relevance to cosmic Reionization. \texttt{dustyGadget} simulations at present predict, in fact, that these systems contribute 80\% of the total M$_{\star}$ in a cosmic volume of 50$\rm h^{-1}$~cMpc at $z\sim 7.5$, decreasing to 40\% at $z\sim 4.5$. Assessing the statistical relevance of these galaxies with observations will be then crucial to characterise their properties and correctly model these environments in future simulations (Venditti et al., in prep.). 

\subsection{Main sequence of galaxy formation}
\label{sec:MS}
The galaxy main sequence (MS) of star formation indicates that there is a strong correlation between the star formation rate (SFR) and the stellar mass ($\rm M_{\star}$) of samples of galaxies observed at a given redshift. The MS encapsulates information on the mechanisms and the efficiency of gas conversion into stars at a fixed redshift; while robustly established at low redshift where large galaxy samples are available, several works suggest that it also holds up to the first couple of Gyr \citep{Speagle+2014}.

\begin{figure*}
\centering
\includegraphics[width=\textwidth]{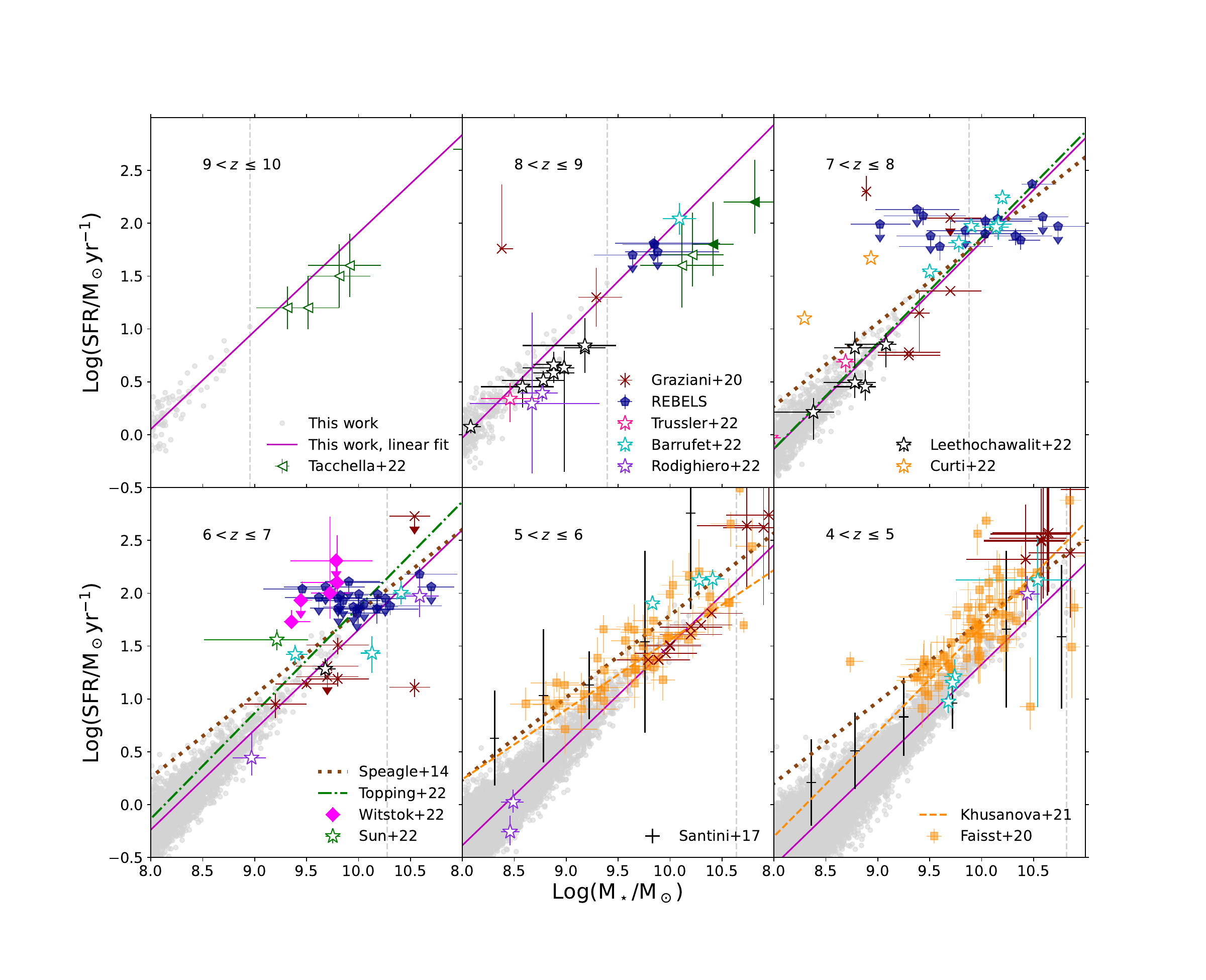}
\caption{MS of star formation in the redshift range between 10 (\textit{top left}) and 4 (\textit{bottom right}): comparison between simulated galaxies  with $\rm{Log(M_{\star}/M_{\odot})} \geq 8$ from U6 (\textit{gray points}), their linear fit in log-scale (\textit{magenta solid line}), and observed galaxies found in the literature. \textit{Green triangles} are the galaxies from the work by \citet{Tacchella+2022} - filled triangles are for the sources that have been spectroscopically confirmed, while empty ones are for those which have not been confirmed yet (see the text for more details), \textit{red crosses} are respectively single observations and upper limits collected in Table 2 of \citet{Graziani+2020}. \textit{Blue pentagons} are the observations obtained by the REBELS survey described in \citet{Topping+2022}, empty stars with \textit{cyan, purple, orange, black, green and pink borders} are the preliminary results from JWST respectively by \citet{Barrufet+2022, Rodighiero+2022, Curti+2022, Leethochawalit2022, Sun+2022, Trussler+2022}. \textit{Magenta diamonds} are the recent observations by \citet{Witstok+2022} and  the \textit{orange squares} show the ALPINE sample by \citet{Faisst+2020} at $4.4 \: < \: z \: < \: 5.9$. Finally \textit{black crosses} are the average values from the analysis of the HST Frontier Field by \citet{Santini+2017}.\textit{Dotted brown lines} are the fitting functions by \citet{Speagle+2014}, \textit{dash-dotted green lines} are from \citet{Schreiber+2015} and the fitting functions of the ALPINE objects obtained by the \citet{Khusanova+2021} are the \textit{orange dashed lines}}.

\label{fig:FigMS}
\end{figure*}

In this section we investigate the redshift evolution of the MS, in the redshift range $4 < z \le 10$, by comparing \texttt{dustyGadget} predictions with samples of isolated galaxies collected in \citet{Graziani+2020} and \citet{Tacchella+2022}, the new datasets offered by the REBELS \citep{Bouwens+2022, Topping+2022} and ALPINE \citep{Faisst+2020,Khusanova+2021} surveys, the analysis of HST Frontier Fields \citep{Santini+2017}, the galaxies observed at $6 < z \leq 7 $ by \citet{Witstok+2022}, and with the recent determinations based on JWST ERO and ERS \citep{Barrufet+2022, Curti+2022, Leethochawalit2022,Rodighiero+2022, Sun+2022, Trussler+2022}. For comparison, we also show extrapolations of the MS based on low redshift observations   (\citealt{Speagle+2014}, dotted lines). 
Figure \ref{fig:FigMS} shows galaxies found in U6 with $\rm{Log(M_{\star}/M_{\odot})} > 8$ as grey dots, while their linear fits are indicated as magenta solid lines. Here and in the following figures, vertical gray dashed lines indicate the maximum stellar mass ($\rm{M_{\star,max}}$) found in the simulation at each redshift, thus, the fit above $\rm{M_{\star,max}}$  has to be interpreted as extrapolation based on $\rm{Log(M_\star/M_\odot) < Log(M_{\star,max}/M_\odot)}$. At $4 < z \le 6$ the galaxy MS has been reliably measured over a large range of stellar masses by the analysis of HST Frontier Fields (\citealt{Santini+2017}, black crosses) and by the ALPINE sample (\citealt{Faisst+2020}, orange squares). Before JWST early results, the available constraints at $z > 6$ were limited at the high-mass end, for galaxies with $\rm{Log(M_{\star}/M_{\odot})} > 9$. When needed, we multiply the stellar masses and star formation rates for the conversion factors by \citet{Madau&Dickinson_2014} to convert them from others IMF to the \citet{Salpeter+1955} one.

The first four panels show the results obtained by \citet{Tacchella+2022} with a sample of 11 bright galaxy candidates\footnote{Note that among these galaxies, previously selected in the CANDELS fields by \citet{Finkelstein+2021}, only three are spectroscopically confirmed so far: EGS-6811, EGS-44164 and GOODSN-35589, respectively at z = 8.68, 8.66 and 10.96 (filled green triangles).} (green empty triangles), together with additional observations of single galaxies collected in \citet{Graziani+2020} (red crosses) and the 5 bright Lyman-break galaxies at $z \sim 7$ by \citet{Witstok+2022} (magenta diamonds). At $z > 6.5$ a very important improvement has been achieved by the REBELS collaboration \citep{Bouwens+2022}, which boosted by a factor of 3 the number of bright ISM-cooling lines discovered, significantly extending the number of available objects observed during the Epoch of Reionization (EoR). Here we show the results obtained by \citet{Topping+2022} (blue pentagons), where M$_{\star}$ are computed using a non-parametric star formation history (SFH). Given the limited range in the stellar mass, \citet{Topping+2022} fixed the slope of the main sequence to the values determined by \citet{Schreiber+2015} at $z = 7$ and constrained the MS normalization for the REBELS sample, finding $\rm{Log(SFR/M_\odot {\rm yr}^{-1})} = \rm{Log(M_\star/M_\odot)} - 8.12$ (see the dash-dotted line in the top-right panel of Figure \ref{fig:FigMS}). 

\begin{figure*}
\centering
\includegraphics[width=\textwidth]{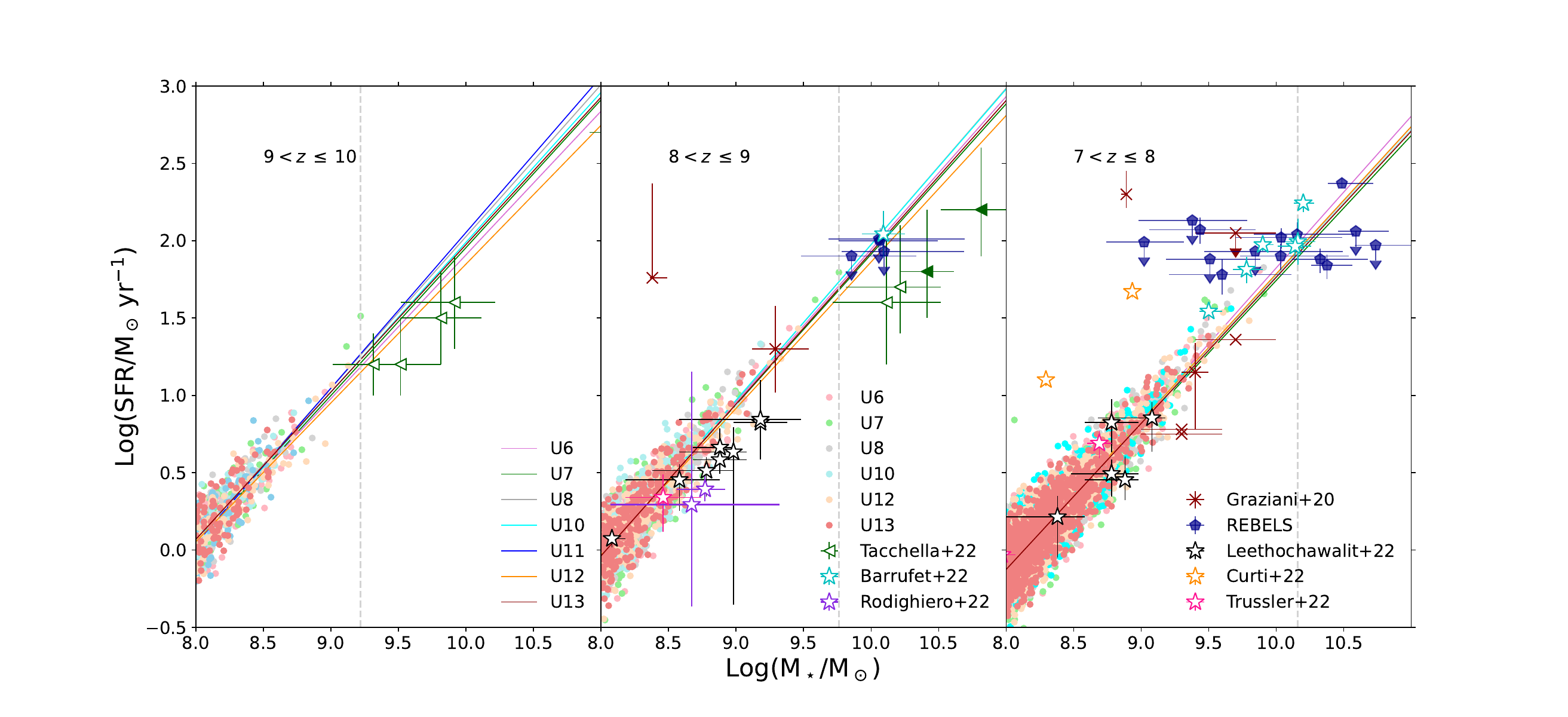}
\caption{Here we explore the variance among our 8 independent simulations by comparing the linear fit of each universe from U6 to U13. The  \textit{solid lines} indicate the linear fit of each universe, each one with a different color, while  the \textit{points} are the simulated galaxies with $M_{\star} \geq 10^8 M_{\odot}$.
For $z > 7$ different universes have different slopes, while at lower $z$ the 8 simulations converge to a common trend.}
\label{fig:FigCosmicV}
\end{figure*}

A collection of data obtained by independent studies based on JWST ERO and ERS is also reported in Figure \ref{fig:FigMS} (empty stars). Although the spectroscopic confirmation is available only for a small number of sources, this figure shows the enormous potential of JWST in constraining the slope of the galaxy MS at $z > 6$. The \texttt{dustyGadget} fit at $7 < z \leq 9$ appears very consistent with some JWST data at the low-mass end \citep{Rodighiero+2022,Trussler+2022, Leethochawalit2022} and, when extrapolated at the high-mass end, with JWST data from \citet{Barrufet+2022} and with the REBELS sample, favouring a steeper slope compared to \citet{Speagle+2014}; it is also in good agreement with the extrapolation of \citet{Schreiber+2015} done by \citet{Topping+2022} (see also Section \ref{sec:MSslopeevo} for the fitting functions based on \texttt{dustyGadget} predictions). While our simulation lacks a significant statistics at the extreme SFR and stellar mass end (see however Figure \ref{fig:FigCosmicV}, where we show the results for all our simulated cosmic volumes), we point out that some studies find a piecewise fit to the star forming sequence at high-redshift \citep{Lovell+2021}, to account for the bending seen in the sequence \citep{Popesso+2022, Sandles+2022}. A comparison with independent model results is illustrated in Figure \ref{fig:FigMS_sim} and a discussion of the evolution of the MS slope is presented in Section \ref{sec:MSslopeevo}.

In the last two redshift panels, we show that our simulated sample is consistent with a number of ALPINE galaxies found in the post-EoR epoch ($6 \: < \: z \: < \: 4$), i.e. we find that $77 \%$ of the simulated candidates in the bottom mid panel and $73 \%$ in the bottom right panel are consistent with the ALPINE galaxies. In these redshift windows, a good agreement is also found with the trend suggested by the estimates of \citet{Santini+2017}, particularly at $4 < z \leq 5$, although the large standard deviations associated with the mid-points of mass bins do not allow to place stringent constraints. JWST early results allows us to constrain both the high-mass end (i.e. $\rm{Log(M_\star/M_\odot)}\geq 9.5$, see for example the data from \citealt{Barrufet+2022} and \citealt{Rodighiero+2022}), as well as the low-mass end \citep{Rodighiero+2022} of the relation.

Finally, we compare the linear fit of our simulated universe (magenta solid line) with the fitting function of the MS by \citet{Speagle+2014} (brown dotted line) and the fitting functions obtained from the ALPINE data by \citet{Khusanova+2021} (dashed orange lines).  In the redshift range $4 < z \leq 6$ our fit seems to predict a steeper MS compared to \citet{Speagle+2014}, while at $4 < z \leq 5$ a shallower slope is found compared to \citet{Khusanova+2021}\footnote{We remind the reader that the results are sensitive to the way the data is binned/aggregated in redshift. For example, in the work done by \citet{Faisst+2020} the ALPINE collaboration finds that the entire galaxy sample $4 < z < 5.9$ (orange squares) is compatible with the relation found by \citet{Speagle+2014} at $z = 5$ within a $\pm 0.3$ dex width.}. In the next section we provide a more in-depth analysis of our results.

\subsection{Additional properties of the galaxy main sequence}
\label{sec:AddMS}
In this section we discuss the impact of simulated cosmic variance and a possible redshift evolution of the slope of our fits. A comparison with the findings of other simulations adopting different simulated cosmic scales and numerical schemes is presented in Appendix \ref{sec:AppMS}.

\subsubsection{Impact of simulated cosmic variance} 
To understand the impact of cosmic variance in our predictions, in Figure \ref{fig:FigCosmicV} we compare the linear fits of the predicted MS across the eight simulated volumes in the same redshift bins of Figure \ref{fig:FigMS}. We limit to the higher three redshift intervals, as at $z \sim 7$ all the predictions tightly converge. Due to fewer statistics, the largest deviation is found at the highest redshifts, with a deviation in the slope of $\Delta \rm{m} = 0.11$, while it reduces to $\Delta \rm{m} = 0.04$ in the lowest redshift bin. Within the above variations, our simulations confirm an even better agreement between the trends followed by simulated and observed galaxies. Note, for example, that in the redshift range $7 < z \leq 10$ a better consistency is found with  the galaxies analyzed in \citet{Tacchella+2022, Graziani+2020,Bouwens+2022,Leethochawalit2022,Curti+2022,Trussler+2022,Barrufet+2022}. As already mentioned, vertical gray dashed lines indicate $\rm{M_{\star,max}}$ found in each simulation. Notice how, once we take into account all the available simulated cubes\footnote{In this work we usually consider U6 as our reference run (RefRun), unless otherwise specified.} , there is an increase in the statistics of the simulated objects resulting in higher upper limits for the stellar masses (i.e. vertical lines move towards higher mass values once compared to those in Figure \ref{fig:FigMS}).

\begin{figure}
\centering
\includegraphics[width=\columnwidth]{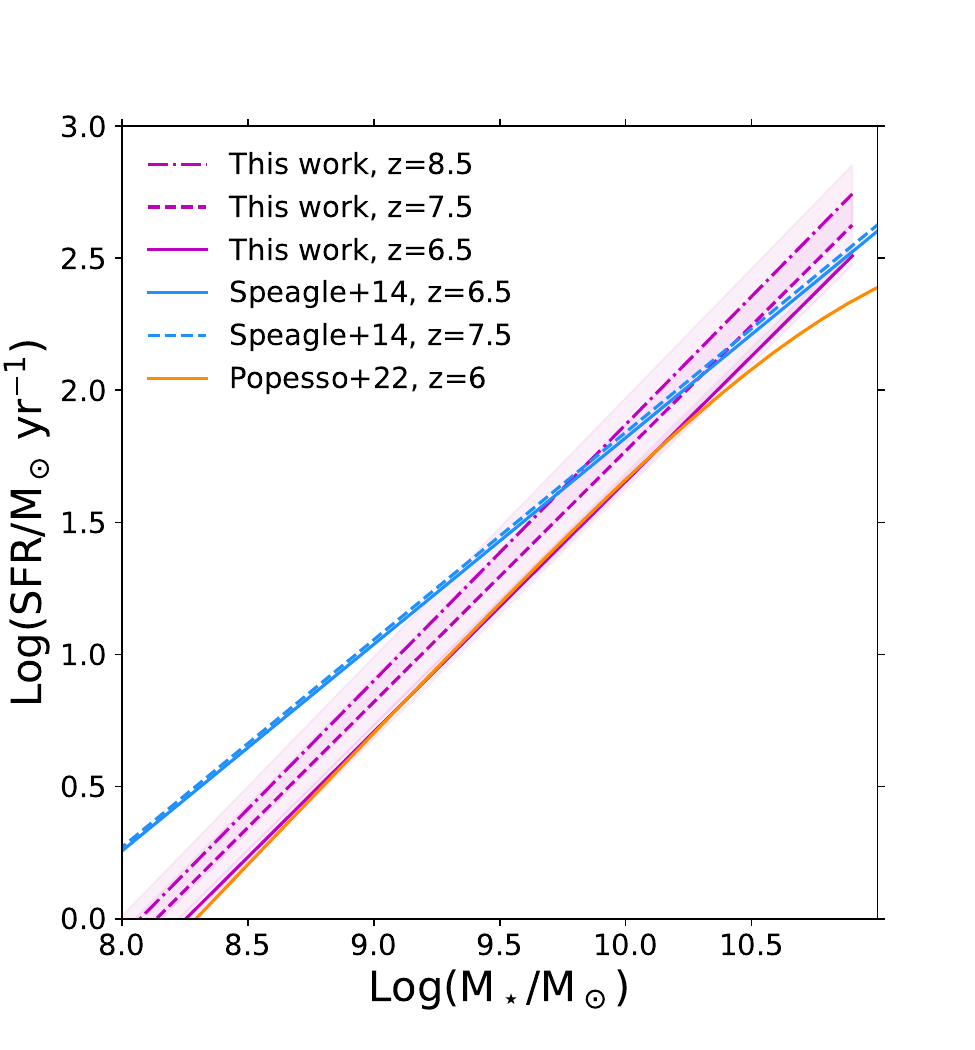}
\caption{Linear fits of the data (\textit{magenta}) from our simulations, considering galaxies with $\rm{Log(M_{\star}/M_{\odot})} \geq 8$. Each fit has been performed in three different redshift bins $6<z\le 7$, $ 7<z\le 8$, $8<z\le 9$ considering all the available cubes at these redshifts, Together with the linear fit we also show here the standard error associated with the angular coefficient (\textit{pink shaded regions}). In \textit{orange, green and light-blue} we report respectively the best fits obtained by \citet{Popesso+2022} at $z=6$, \citet{Schreiber+2015} at $z=7$ and \citet{Speagle+2014} at $z=6.5$ and $z=7.5$.}
\label{fig:FigMS_slope}
\end{figure}

\subsubsection{Redshift evolution of the main sequence} 
\label{sec:MSslopeevo}
There is a general consensus about the increasing normalization of the MS with redshift, which is associated with a higher rate of gas accretion onto galaxies in the early Universe. However, a possible evolution in the slope of the MS is hard to constrain because of its dependence on the sample selection and the SFR tracer adopted \citep{Speagle+2014}. At $z<4$, observations seem to suggest that the MS is characterized by a constant slope that is close to unity when considering $\rm{M_{\star}}<10^{10.5} \rm{M_{\odot}}$ \citep{Whitaker+2014,Tasca+2015,Schreiber+2015,Tomczak+2016,Santini+2017}, suggesting a similarity in the gas accretion histories of galaxies. At higher redshifts \citet{Khusanova+2021} investigated the MS using the ALPINE sample, finding no evidence for a change in the MS slope between $z\sim 4.5$ and $z\sim 5.5$. Very recently \citet{Popesso+2022,Daddi+2022} investigated the MS evolution in the redshift ranges $0<z<6$ and $0<z<4$ respectively. In both cases, they find that at the faint-end the MS has a linear slope that does not change with time, while at large stellar masses the MS bends, with a turn-over mass that is evolving with time. They interpret this result as an indication of a transition between a regime where star formation is efficiently sustained by gas accretion to a regime where star formation is suppressed by the interplay between the hot gas in massive halos and central black hole feedback \citep{Popesso+2022}. However their turn-over mass is $\rm{Log(M_\star/M_\odot)} \sim 10.9$ at $z \sim 6$ and $\rm{Log(M_\star/M_\odot)} \sim 10.6$ at $z \sim 4$. As it can be seen from the two bottom right panels of Fig. \ref{fig:FigMS}, our simulation predicts zero or very few galaxies with masses above the turn-over mass and, as a consequence, our fit is sensitive to lower mass objects. Because of this, our simulation can only sample the regime of stellar masses where observations do not expect a significant evolution in the MS slope and galaxies are found to evolve with a constant SFR per unit stellar mass.

In order to study the main sequence slope at higher redshift ($z>6$) and to have a significant number of simulated candidates, we restrict our analysis to the redshift range $6 <z \le 9$ to explore if our simulations predict any evolution in the MS slope. Figure \ref{fig:FigMS_slope} shows the linear fit (green) we performed on our simulations when considering galaxies with $\rm{Log(M_{\star}/M_{\odot})} \geq 8$ and the standard error associated with the value of the angular coefficient at each $z$ (shaded region). In particular, solid lines correspond to the linear fit in the redshift range $8<z\le 9$, dashed lines to $7<z\le 8$ and dashed-dotted lines to $6<z\le 7$. 

The best fits we obtained for each of the following redshift bins $6<z\le 7$, $ 7<z\le 8$, $8<z\le 9$ are respectively:
\begin{gather}
    {\rm Log \: SFR} =  (0.948 \pm 0.004){\rm Log \:M_{\star}}-(7.83 \pm 0.03) \\
    {\rm Log \: SFR} =  (0.951 \pm 0.007){\rm Log \: M_{\star}}-(7.73 \pm 0.06)  \\
    {\rm Log \: SFR} =  (0.98 \pm 0.01){\rm Log \: M_{\star}}-(7.8 \pm 0.1)
\end{gather}
where $\rm{Log \: SFR = Log(SFR/M_{\odot} yr^{-1})}$ and $\rm{Log \: M_{\star}= Log(M_{\star}/M_{\odot})}$. 

Both slope and normalization of the simulated MS appear to be constant, within the errors. 

In Figure \ref{fig:FigMS_slope} we also compare our results with the best fit functions by \citet{Popesso+2022} and \citet{Speagle+2014} respectively in orange and blue. Both relations are calibrated on lower redshifts observations ($0<z<6$ and $0<z<4$) but they can be exploited to assess the MS evolution at higher $z$. 

As expected, we are not able to reproduce the bending found by \citet{Popesso+2022}, whereas our predicted slope at $6 < z \leq 9$ is steeper than the extrapolation of \citet{Speagle+2014} and favoured by JWST early results, as discussed above.

\begin{figure*}
\centering
\includegraphics[width=\textwidth]{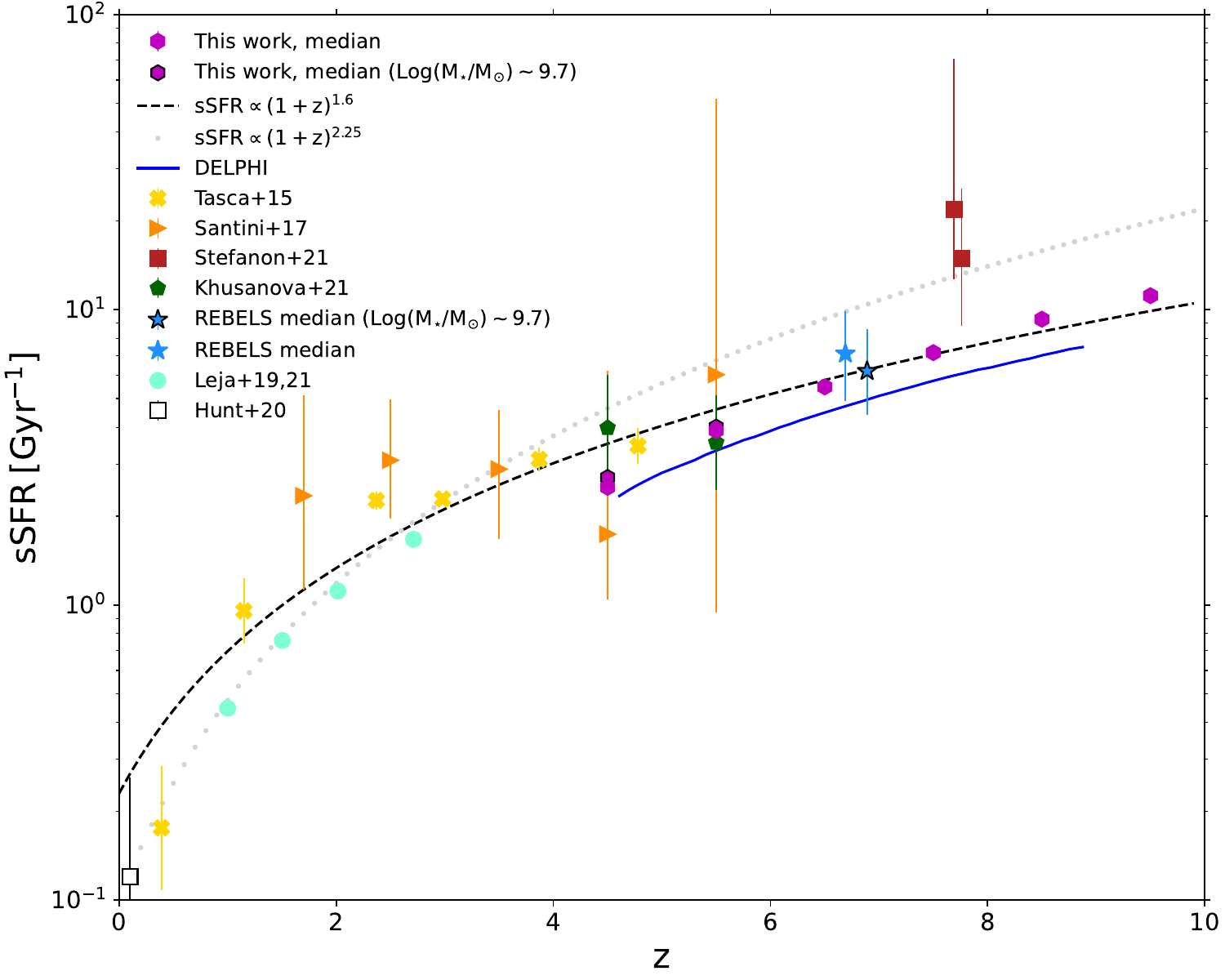}
 \caption{The redshift evolution of the sSFR, we show in \textit{magenta} the results from our simulations in the redshift range $z=4-10$ - the statistical errors for our data are smaller than the symbols - and in the mass range $8 < \rm{Log(M_{\star}/M_{\odot})} < 11$. In \textit{magenta with black borders} we show the sSFR in the mass range $9.6 < \rm{Log(M_{\star}/M_{\odot})} < 9.8$, compatible with the median mass value of $\rm{Log(M_{\star}/M_{\odot})} \sim 9.7$ present in observed samples. We compare our results with the models by \citet{Dekel+2009, Dave+2011a, Sparre+2015} (\textit{gray dotted line}), \citet{Topping+2022} (\textit{black dashed line}) and \citet{Dayal+2022} (\textit{blue solid line}), and with the observations by the REBELS survey \citet{Topping+2022} (\textit{blue stars}), \citet{Tasca+2015} (\textit{yellow squares}), \citet{Santini+2017} (\textit{orange triangles}), \citet{Stefanon+2021a} (\textit{red squares}), \citet{Khusanova+2021} (\textit{green pentagons}), \citet{Leja+2021} (\textit{turquoise points}) and the \textit{empty square with black border} is the estimate of the sSFR at z=0 by \citet{Hunt2020}}.
\label{fig:FigsSFR}
\end{figure*}

\subsection{Redshift evolution of the specific star formation rate}
\label{sec:sSFR}
The specific SFR (sSFR), i.e. the SFR per unit stellar mass, is often used as an additional diagnostic of how SFR and M$_{\star}$ are related. Observational determinations of the sSFR at high-$z$ have largely benefit from improved constraints on dust-obscured star formation coming from dust continuum detections with ALMA \citep{Khusanova+2021, Topping+2022}. This section investigates the redshift evolution of the median values of the sSFR in our simulations\footnote{The sSFR is derived accounting for galaxies with masses in the range $\rm{ Log(M_{\star}/M_{\odot}) =[8-11]}$.}. Figure \ref{fig:FigsSFR}, (adapted from \citealt{Topping+2022}), shows \texttt{dustyGadget} predictions (magenta hexagons) and its comparison with theoretical models and observations. 
Theoretical models in which the growth of galaxies is mainly regulated by gas accretion through cold streams \citep{Dekel+2009, Dave+2011a, Sparre+2015}, predict a  sSFR rapidly rising toward higher redshifts, with a dependence proportional to $(1+z)^{2.25}$ (gray dotted line). Deviations from this estimate could arise from a different behaviour of feedback processes in the high-z Universe. For example, a different efficiency in gas accretion could significantly alter the rate of star formation, and JWST data will certainly provide clues on how fast gas is converted into stars within the EoR.

At lower redshift, larger samples of  galaxies with highly reliable spectroscopic redshift are available from: (i) the VIMOS Ultra-Deep Survey (VUDS) (yellow crosses, \citealt{Tasca+2015}), (ii) the deep COSMOS-2015 and 3D-HST rest frame UV-IR photometric catalogs (turquoise points, \citealt{Leja+2019,Leja+2021}), clearly showing a redshift dependence of the sSFR.

The above evolution, on the other hand, is not confirmed by all data, especially at the highest redshifts. Observations from the first four HST Frontier Field clusters (orange triangles) collected with a fixed mass bin $9.5 \leq \rm{Log(M_{\star}/M_{\odot})} \leq 10$ and an average value of $\rm{Log(M_{\star}/M_{\odot})} \sim 9.7$ \citep{Santini+2017} confirm a mild evolution up to $z \sim 6$, while the ALPINE sub-sample analyzed by \citet{Khusanova+2021} (green pentagons), with candidates in the mass bin $9.6 \leq \rm{Log(M_{\star}/ M_{\odot})} \leq 9.8$, shows little to no evolution at $4.5 \leq z \leq 5.5$. The REBELS collaboration (\citealt{Topping+2022}) provided a power-law fitting of the REBELS galaxies and other available measurements adopting a fixed mass bin of $\rm{Log(M_\star/M_\odot)} = 9.6 - 9.8$ and a constant SFH, finding that the sSFR increases with redshift $\propto (1+z)^{1.7 \pm 0.3}$ over the redshift range $z \sim 1 - 7$. Using a non-parametric SFH for REBELS galaxies, which significantly affects the stellar mass derivation at the low-mass end, the evolution at a fixed stellar mass of $\rm{ Log(M_{\star}/M_{\odot})} = 9.7$ is only mildly affected\footnote{The difference between stellar masses derived assuming a constant SFH or a non-parametric SFH is particularly significant for young and low-mass galaxies (i.e. $\rm{Age < 10 \, Myr}$ and $\rm{Log(M_\star/M_\odot)} < 9$ respectively). A detailed comparison of stellar masses derived by the two methods applied to the REBELS sample can be found in \citet{Topping+2022}.}, with sSFR $\propto (1+z)^{1.6 \pm 0.3}$ in the same redshift range (black dashed line). Note, however, that - among the low-redshift observations - only the \citet{Leja+2019, Leja+2021} derivations are based on a non-parametric SFH, and this would favor a steeper evolution with redshift.  

A non-parametric SFH increases the M$_{\star}$, reducing the sSFRs to $\rm sSFR=7.1^{+2.8}_{-2.2} \: \rm{Gyr^{-1}}$ when considering the entire REBELS sample, or to  $\rm sSFR=6.2^{+2.4}_{-1.8} \: \rm{Gyr^{-1}}$ when considering the mass range $9.6 <\rm{Log(M_{\star}/ M_{\odot})} < 9.8$, consistent with the ALPINE sample, at $z\sim 7$ \citep{Topping+2022}. Finally, \citet{Stefanon+2021a} (red squares) derived the sSFRs for a sample of Lyman-Break galaxies at $z\sim 8$ with $\rm{M_{UV}}$ similar to that of REBELS showing even higher sSFR at $z \sim 8$, compatible with the original trend (gray dotted line).   

Recently, semi-analytic models provided theoretical estimates of the sSFR predicted by the REBELS collaboration (see \citealt{Topping+2022}) as shown in Figure \ref{fig:FigsSFR}. The \texttt{DELPHI} model \citep{Dayal+2014,Dayal+2022} for example, predicts a power-law evolution at $z \geq 4$ (solid blue line), consistent with REBELS power-law fitting up to a normalization factor. \texttt{dustyGadget} predictions are in agreement with the smooth evolution (black dashed) proposed by \citet{Topping+2022}. However, to guarantee consistency across observed and simulated data samples, a uniform stellar mass range is required. For the above reason we first computed the sSFR on the full galactic sample discussed in this paper ($\rm{Log(M_\star/M_\odot)} \geq 8$) (magenta hexagons), and then we made the same estimates considering the mass range $9.6 < \rm{Log(M_{\star}/M_{\odot})} < 9.8$, compatible with the median mass value of $\rm{Log(M_{\star}/M_{\odot})} \sim 9.7$ present in observed samples. These results are shown in Figure \ref{fig:FigsSFR} as magenta hexagons with black borders. More specifically, we find sSFR = 2.5 Gyr$^{-1}$ at $z = 4.5$ and 3.9 Gyr$^{-1}$ at $z=5.5$, respectively, where a sufficiently large number of simulated galaxies is available. When restricting to the aforementioned stellar mass bin, our results increase by $\sim 3 - 8\%$ and we find sSFR = 2.7 Gyr$^{-1}$ at $z = 4.5$ and 4.0 Gyr$^{-1}$ at $z=5.5$. Overall, our simulation suggests an increase of the sSFR with redshift in the range $4 < z < 10$, consistent with current observations and with theoretical expectations based on increased baryon accretion rates at high-redshifts. As a comparison with the Local Universe, in Fig. \ref{fig:FigsSFR} we also show the sSFR at $z = 0$ estimated by \citet{Hunt2020}\footnote{In estimating the sSFR we considered the mass range $8 < \rm{Log(M_\star/M_\odot)} < 11$ in order to be consistent with the one analysed in the present work. Notice that even at $z=0$, where the data sample is larger, the sSFR changes by $\sim 0.5$~dex depending on the mass range considered.}. Assuming the mid point of the observed mass range to be Log$(\rm{M_\star/M_\odot}) = 9.5$ and using Equation 1 of \citet{Hunt2020}, we obtain a sSFR of $0.12 \: \rm{Gyr^{-1}}$.

\subsection{Halo-Stellar mass relation}
\label{sec:halo-stellar}

\begin{figure*}
\centering
\includegraphics[width=\textwidth]{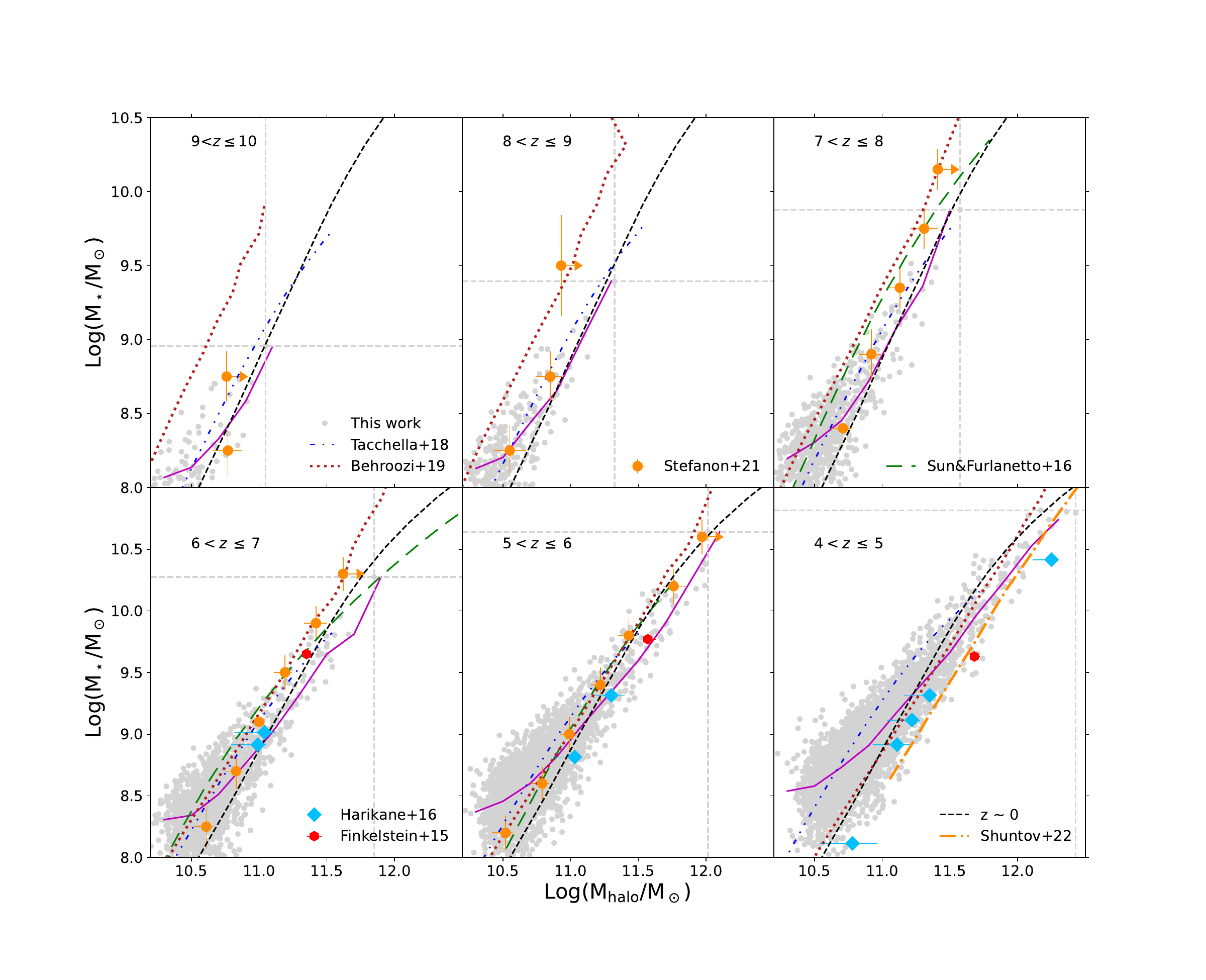}
\caption{Redshift evolution of the stellar mass ($\rm{M_{\star}}$) and halo mass ($\rm{M_{halo}}$) relation. Here we compare our simulated galaxies (\textit{gray points}) and their median trends (magenta solid lines)} with the observations by \citet{Stefanon+2021} (\textit{orange circles}), \citet{Harikane+2016} (\textit{light blue diamonds}), \citet{Finkelstein+2015} (\textit{red hexagons}), \citet{Shuntov+2022} (\textit{dash-dotted orange line}) and the predictions obtained by \citet{Behroozi_2019, Tacchella+2018, SunFurlanetto+2016} (respectively in \textit{red dotted lines, dash double dotted blue lines and loosely dashed green lines}). The behaviour of the $\rm{M_{halo}-M_\star}$ relation at $z \sim 0$ \citep{Behroozi_2019} (\textit{dashed black lines}) has been shown to guide the eye.

\label{fig:FigMhalo-Mstar}
\end{figure*}
The relation between the dark matter mass of a halo ($\rm{M_{halo}}$) and the stellar mass of its galaxies is often assumed by semi-analytic or data-constrained models which do not explicitly model baryonic processes with a hydrodynamical approach. Accordingly to these theoretical schemes \citet{Rees+1977, White+1978, Fall+1980}, the assembly of the stellar mass is driven by the large-scale process of dark matter accretion either through mergers or smooth accretion from filaments, prompting the flow of cold gas into the central galaxy. As a result, the stellar-to-halo mass relation (SHMR) is a proxy of the star formation efficiency. The same scaling relation is often adopted by observers to provide lower limit constraints on DM overdensities associated with luminous objects. We multiply by a factor 1.7 \citep{Madau&Dickinson_2014} the stellar masses to convert them from a \citet{Chabrier+2003} to a \citet{Salpeter+1955} IMF.

Figure \ref{fig:FigMhalo-Mstar} shows results from our simulations (gray dots) and their median trends (magenta solid lines). Vertical and horizontal dashed lines show respectively the maximum halo mass ($\rm{M_{halo,max}}$) and $\rm{M_{\star,max}}$ found in the RefRun. We also report recent observational constraints on the SHMR from abundance matching techniques by \citet{Stefanon+2021} (orange circles) and \citet{Finkelstein+2015} (red hexagons). Estimates by \citet{Harikane+2016}, which rely on the two-point correlation function of LBGs, are shown as light blue diamonds, while the recent estimates based on the COSMOS2020 catalog from \citet{Shuntov+2022} are in orange dash-dotted lines. We also show the redshift dependent SHMR predicted by the data-constrained model of \citet{Behroozi_2019}\footnote{For this comparison, we considered the average halo masses as a function of observed stellar masses found in the database of \citet{Behroozi_2019}.} (red dotted lines and shaded areas) and additional theoretical models which either assume a constant SHMR above $z > 4$ \citep{Tacchella+2018}, or introduce a redshift dependent conversion efficiency between the halo accretion rate and the star formation rate (\citealt{Moster+2018}, dashed-dotted light blue lines, \citealt{SunFurlanetto+2016}, loosely dashed green lines). Finally, as a reference to predictions in the Local Universe, we also show the behaviour of the SHMR relation at $z \sim 0$ by \citet{Behroozi_2019}. 

At all redshifts, our simulations indicate the expected monotonic increase of $\rm{M_\star}$ with $\rm{M_{halo}}$. The simulated systems show a significant scatter in the relation, particularly at the low-mass end, where we have larger statistics. At $5 < z \leq 6$, DM halos with mass Log$(\rm{M_{halo}/M_\odot}) = 10.5$ are predicted to host galaxies with stellar masses $8 \leq \rm{Log(M_\star/M_\odot)} < 9$, likely reflecting the ongoing process of galaxy assembly and the large variety of SFHs experienced by these systems. When compared to other models and observations, our median trends are generally consistent with previous results. 
 
However, at $z \leq 6$ our simulated galaxies appear to grow in mass more efficiently, with a deviation in the median trend at $\rm{ Log( M_{\rm h}/M_\odot)} < 11$ that progressively increases with time. This is consistent with the estimated low-mass end of the SMF, which appears to predict a larger number density of systems compared to some observational determinations (see Section \ref{sec:StellarMF}) and model predictions (see Appendix A) at $z \le 6$, while being in agreement with other observational and theoretical studies. This may suggest that in low-mass galaxies feedback may be more effective than modelled by \texttt{dustyGadget} (see for example \citealt{Graziani2015, Graziani2017}), despite the  encouraging agreement between our predicted galaxy MS and JWST early results in the mass range $10^8-10^{11} \: \rm{M_\odot}$ at $z=4-9$.

\subsection{Dust-to-stellar mass scaling relation}
\label{sec:dust-star}
\begin{figure*}
\centering
\includegraphics[width=\textwidth]{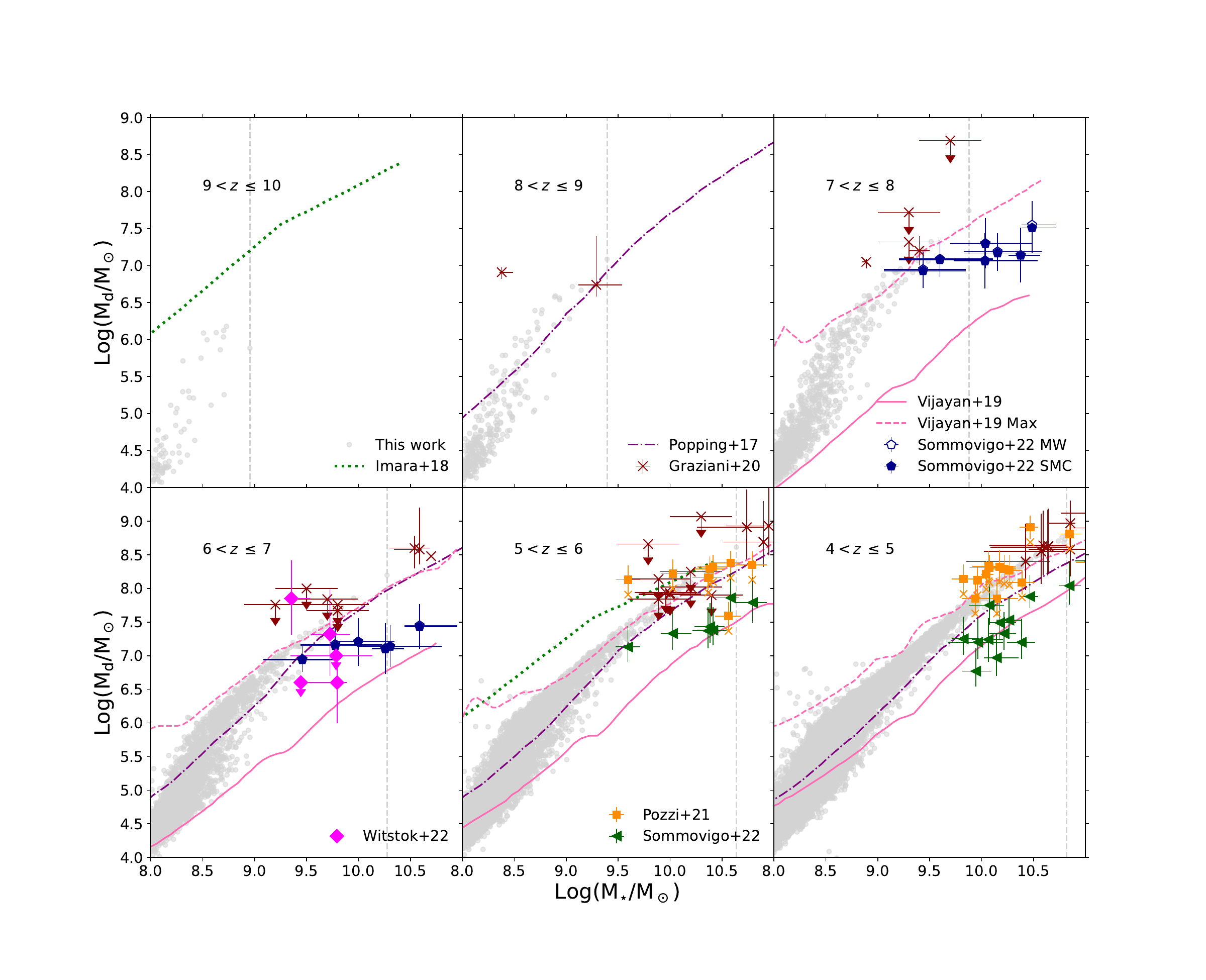}
\caption{Dust mass ($\rm{M_d}$) as a function of the stellar mass ($\rm{M_{\star}}$) both in units of $\rm{M_{\odot}}$ for simulated galaxies with Log$(\rm{M_{\star}/M_{\odot}})>8$ extracted from simulation U6 (gray points). The evolution is investigated in the redshift range $4<z \le 10$. \textit{Red crosses} are observations from the compilation by \citet{Graziani+2020} (see their Table 2), \textit{blue pentagons} are from the REBELS survey \citet{Sommovigo+2022} and \textit{magenta diamonds} are observations from \citet{Witstok+2022}.\textit{Orange squares and crosses} are the dust masses obtained within the ALPINE survey by \citet{Pozzi+2021} assuming respectively $\rm{T_d} = 25\:\rm{K}$ and $35\:\rm{K}$, while \textit{green triangles} are the recent estimates, for some of the ALPINE galaxies, by \citet{Sommovigo2022alpine} where $\rm{<T_d>} = 48 \pm 8 \: \rm{K}$. We also compare our results with other independent studies: \textit{dash-dotted violet lines} are the averaged trends computed by \citet{Popping+2017}, \textit{pink solid and dashed lines} are respectively the fiducial and maximum models of \citet{Vijayan+2019}, and \textit{green dotted lines} show predictions by \citet{Imara+2018}.}
\label{fig:FigMdust-Mstar}
\end{figure*}

This section updates our predictions on the relation between dust ($\rm{M_d}$) and stellar mass content in high-z galaxies. In \citet{Graziani+2020} the same relation was shown for the first time as predicted by a lower mass resolution simulation performed on a $30\rm{h}^{-1}$~cMpc cosmological volume. The highest mass resolution of the new simulations allows to (i) better model processes occurring in the ISM of galaxies with Log$(\rm{M_{\star}/M_{\odot}}) \geq 9$, (ii) investigate in more detail systems in the intermediate mass range ($8 \leq {\rm Log}(\rm{M_{\star}/M_{\odot}}) \leq 9$), and (iii) collect a larger number of assembly histories from the eight simulations, thanks to the increased statistics, particularly at the high-mass end. Here the simulations are also compared with the REBELS sample (dust mass estimates provided by \citealt{Sommovigo+2022}), with ALPINE galaxies (estimated in \citealt{Pozzi+2021} and \citealt{Sommovigo2022alpine}) and with a recent dataset provided in \citet{Witstok+2022}. Also in this case, we multiply by a conversion factor 1.7 the stellar masses to convert them from a \citet{Chabrier+2003} to a \citet{Salpeter+1955} IMF.

Figure \ref{fig:FigMdust-Mstar} shows the redshift evolution of the $\rm{M_d} - \rm{M_\star}$ relation for galaxies with $\rm{Log(M_{\star}/M_{\odot}}) \geq 8$ found in our RefRun (U6, gray points), while vertical dashed lines show the $\rm{M_{\star,max}}$ value at each redshift. We also show predictions from the empirical model by \citet{Imara+2018} (green dotted lines), the semi-analytic model of \citet{Popping+2017} (purple dash-dotted lines), and the median/maximal relation of \citet{Vijayan+2019} (pink solid/dashed lines, respectively). 

The new simulation enforces the s-shape trend found in \citet{Graziani+2020} with increased statistics, extending it towards larger stellar masses. The $\rm{M_d - M_\star}$ relation that we find confirms the good agreement with the predictions by \citet{Popping+2017} in the redshift range $4 < z \leq 9$, and lies in between the median and the maximum relations of \citet{Vijayan+2019}. However, we systematically predict, both at $9 < z \leq 10$ and at $5 < z \leq 6$, less dust-enriched systems compared to \citet{Imara+2018}.

The observational dataset relies on single dusty galaxies (including the sample of ALESS galaxies) collected in \citet{Graziani+2020} (red, Table 2 of that paper), on ALPINE continuum detection (orange, \citealt{Pozzi+2021})\, on the recent estimates of the dust mass budget in the ALPINE sample by \citet{Sommovigo2022alpine} (green) and on dust masses derived from dust continuum detection of REBELS galaxies \citep{Inami2022} by \citet{Sommovigo+2022} (blue) assuming a Small Magellanic Cloud (SMC) dust model\footnote{Similarly to what we showed in the previous sections, the stellar masses considered here for REBELS galaxies are computed using a non parametric SFH as in \citet{Topping+2022}.}. 

Dust mass estimates strongly depend on the assumed cold dust temperature: \citet{Pozzi+2021} investigated this dependence for the ALPINE sample, finding that going from T$_{\rm d}=25 $~K (fiducial value) to T$_{\rm d}=35$~K results in a decrease of the dust mass by $60\%$ (these estimates are respectively shown as orange squares and crosses in Figure  \ref{fig:FigMdust-Mstar}). Using the method described in \citet{Sommovigo+2021}, very recently \citet{Sommovigo2022alpine} derived new dust mass estimates of some of the ALPINE galaxies already analysed by \citet{Pozzi+2021}. The new analysis leads to warmer dust temperatures $\rm{<T_d>} = 48 \pm 8 \: \rm{K}$ and, as a consequence, to ${\rm M_{d}}$ up to 7 times lower than those previously reported. The same analysis technique has been applied to a subsample of REBELS galaxies with [CII] and dust-continuum detections \citep{Sommovigo+2022}, finding a $\rm{T_d}$ that varies in the range between $39-58$~K, with an average dust temperature of $<\rm{T_d}> = 47 \pm 6$~K. The red circles (ALESS galaxies from Table 2 of \citealt{Graziani+2020}) are estimates with a $\rm{T_d}$ around $45$~K, while the dust mass upper limits, since the dust temperature anti-correlates with the dust mass, are estimated using $\rm{T_d} \sim 25$~K. As in \citet{Graziani+2020}, the gray points of simulated galaxies are in good agreement with estimates or compatible with upper limits of M$_{\rm d}$ obtained for singly detected objects. Despite the large uncertainties in the M$_{\star}$ of galaxies observed in the EoR, we are in agreement with REBELS galaxies with masses $\rm Log(\rm M_{\star}/\rm M_{\odot}) \lesssim 9.5$, while objects with Log(M$_{\star}/\rm M_{\odot}) > 9.5$ have estimated dust masses systematically lower than our predictions, indicating that either our high-mass objects are too dusty or that the dust temperature for some of these sources may have been over-estimated. Radiative transfer simulations performed with \texttt{SKIRT} (\citealt{Baes+2015}) on \texttt{dustyGadget} simulated galaxies, and a close comparison with photometric properties of the REBELS sample will help to shed some light on the above discrepancy (Schneider et al., in prep.). Interestingly enough, we find that at lower redshift the simulated sample is globally compatible with ALPINE/ALESS estimates found in \citet{Pozzi+2021} and \citet{Graziani+2020} for galaxies with Log$(\rm{M_{\star}/ M_{\odot}}) > 9.5$, with only a few exceptions that lie above or below the trend followed by the simulated systems. Conversely, when compared to dust masses derived by \citet{Sommovigo2022alpine} for a subsample of ALPINE systems, our simulation predicts higher dust masses.

A further comparison with \citet{Witstok+2022}, provides us with five more observed galaxies in the redshift interval $6 < z \leq 7$. These galaxies have stellar masses in the range Log $\rm (M_\star/ \rm M_\odot) = 9.1 - 9.9$ and their FIR SED fits favour the following dust temperatures:  $\rm{T_d}=59^{+41}_{-20}$~K for UVISTA-Z-001, $\rm{T_d}=47^{+40}_{-17}$~K for UVISTA-Z-019, and the extremely low value of $\rm{T_d} = 29^{+9}_{-5}$~K for COS-3018555981. For the two additional sources, dust continuum was not confidently detected in any ALMA band, and they assumed a $\rm{T_d}=50$~K. The resulting dust masses are shown as magenta dots. With the exception of COS-3018555981 for which the low dust temperature favoured by the FIR fit suggests a very high dust mass, all the other sources appear to be consistent with the simulated galaxies, at least within the errorbars. When interpreted at face value, the fact that COS-3018555981 is well above the simulated galaxies may imply a very efficient dust production mechanism in this system \citep{Witstok+2022}, beyond what is predicted by \texttt{dustyGadget} for galaxies of comparable stellar mass and redshift when accounting for stellar dust production and ISM grain growth. 

\begin{figure}
\centering
\includegraphics[width=\columnwidth]{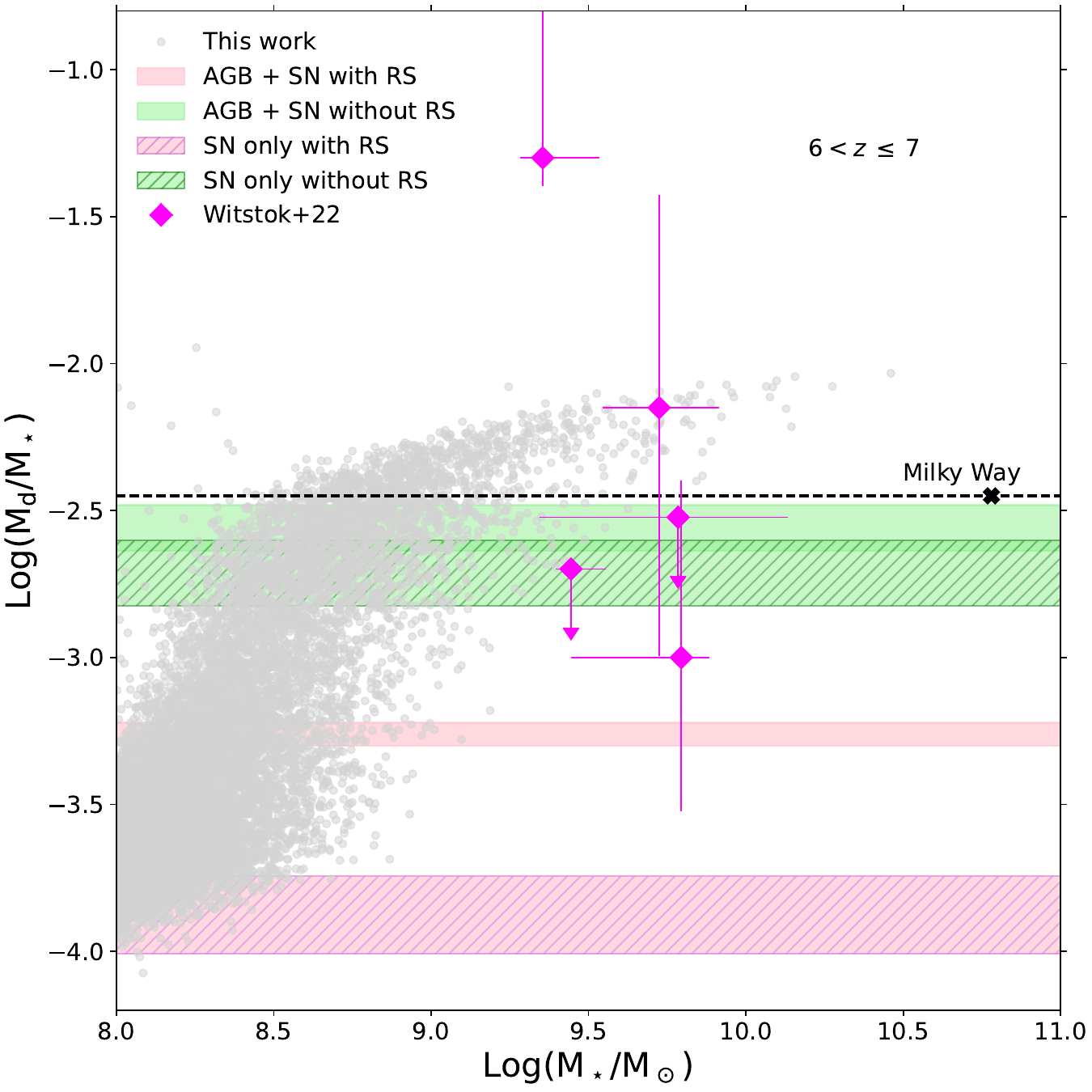}
\caption{The dust-to-stellar mass ratio in the redshift range $6 < z \leq 7$ for the simulated galaxies (\textit{gray dots}) in the RefRun (U6) and in simulations U7, U12 and U13. We also show the values reported by \citet{Witstok+2022} as \textit{magenta points}. The horizontal solid bands represent the value expected for a population age $\rm{t_\star} = 650$ Myr (see the text on how this value has been computed), with upper and lower bounds corresponding to assuming $\rm{Z_\star} = 1 \, \rm{Z_\odot}$ and $\rm{Z_\star} = 0$. While the horizontal hatched bands represent the yields assuming a stellar population age of 15 (18.7) Myr, corresponding to the age of a minimum star that evolves as a core-collapse SN for a stellar metallicity of $\rm{Z_\star} = 1 \, \rm{Z_\odot}$, upper bound ($\rm{Z_\star} = 0$, lower bound). Here we assume the same dust yields adopted in the present \texttt{dustyGadget} simulations with (\textit{pink}) and without (\textit{green}) the SN reverse shock (RS) destruction. For comparison, we also report the dust-to-stellar mass ratio for the Milky Way (\textit{dashed black line} and \textit{black cross}).}
\label{fig:MdustMstarRatio}
\end{figure}

The dust-to-stellar mass ratio can give us some hints on how much dust per unit stellar mass survives the various destruction processes in galaxies. Also, it is a useful quantity to study the evolution of different types of galaxies (see for example \citealt{Calura+2017}). Figure \ref{fig:MdustMstarRatio} shows the $\rm{M_d/M_\star}$ ratio as a function of stellar mass for the simulated galaxies from 4 independent simulated volumes (U6, U7, U12 and U13) at $6 < z \leq 7$. The inferred ratios from observational data points of \citet{Witstok+2022} presented in the bottom left panel of Figure \ref{fig:FigMdust-Mstar} are shown again, together with the estimated Milky Way value (dashed black line and black filled point, \citealt{GrazianiGAMESH2017, GinolfiGAMESH2018}). We also show the yields expected for stellar sources assuming a maximum population age of 650 Myr (computed as the difference between the Hubble time at $z = 6.85$, the measured redshift of COS-3018555981, and the Hubble time at $z = 25$, assumed to be the onset redshift of star formation). Here we assume the same dust yields implemented in the current simulations (see Section \ref{sec:galForm}) and a maximally efficient SN dust production, assuming no reverse shock (RS) destruction. For each of these two cases, represented respectively by pink and green horizontal shaded bands, the minimum (maximum) value corresponds to assuming a fixed stellar metallicity of $\rm{Z_\star} = 0$ ($\rm{Z_\star} = 1 \, \rm{Z_\odot}$). To appreciate the contribution of SNe to early dust production, we also show the same predictions but assuming a stellar population age of 15 (18.7) Myr, which corresponds to the age of the progenitor star with the minimum mass that evolves as a core-collapse SN ($12 \, \rm{M_\odot}$ in our chemical evolution model) for a stellar metallicity of $\rm{Z_\star} = 0$ ($\rm{Z_\star} = 1 \, \rm{Z_\odot}$). These two additional cases are shown as horizontal hatched bands, with the same colour coding of the previous two cases. For the same set of yields (pink and green areas), the difference between the dust-to-stellar mass ratio for a $\rm{t_\star} = 650$~Myr population and the latter cases is due to the contribution of the most massive AGB stars (with masses between $\sim 2 - 2.4 \: \rm{M_\odot}$ and $\sim 8\: \rm{M_\odot}$). 

The comparison between the dust yields and the simulated systems shows that the dust content of each galaxy is the result of a complex interplay between dust production/destruction mechanisms and that grain growth in the ISM contributes to the enrichment of high-mass galaxies \citep{Valiante2014,Mancini2015, Graziani+2020}. While the other sources reported by \citet{Witstok+2022} are consistent with this picture, COS-3018555981 stands out, requiring a substantially more efficient dust production mechanism, an upward revision of the estimated dust temperature, a higher dust emissivity, or a mix of the above. 

\section{Conclusions}
\label{sec:Conclusions}
The present work investigates the build-up of the stellar mass of galaxies at $z \geq 4$ and the scaling relations of their integrated physical properties. We selected observationally well established correlations in the Local Universe, and thanks to a wealth of new data provided by recent high-redshift ALMA Large Programs, such as REBELS \citep{Bouwens+2022} and ALPINE \citep{Faisst+2020}, we were able to benchmark numerical predictions of the \texttt{dustyGadget} model \citep{Graziani+2020} with updated observations, including some of the early release observations of JWST.

With this aim in mind, we performed a new set of eight statistically independent cosmological simulations on a scale of 50$\rm{h^{-1}}$~cMpc in order to increase the statistical sample of predicted galaxies, to account for a larger scatter in their predicted properties and to have access to a wider sample of dusty environments produced by stellar feedback, dynamical encounters, and mergers as well as hydrodynamical effects. The resulting integrated dataset provides a statistically robust sample of dusty galaxies in the stellar mass range  $8.0 \leq \rm Log(M_\star/M_\odot) < 11.0$ suitable to investigate the build-up of stellar mass and the redshift evolution of some  galaxy scaling relations at $z \geq 4$. In particular, we find that:

\begin{itemize}
    \item the total stellar build-up, both in terms of total star formation rate and total stellar mass density, rapidly increases from the onset of star formation occurring around $z \sim 20$, down to $z \sim 4$ with a remarkable agreement with available observations, including JWST ERO and ERS at $z \geq 8$; 
    \item at $4 < z \leq 9$ the stellar mass function predicted by the simulation shows a broad agreement with observations and with independent theoretical predictions. At $z < 8$, \texttt{dustyGadget} predicts fewer massive objects compared to observed samples, due to the limited statistics of massive systems in the simulated volume at these redshifts.
    \item at $7 < z \leq 9$ we find that the simulated galaxy main sequence is in very good agreement with available data, including some of the first JWST ERO and ERS which extends to lower stellar masses the observational constraints placed by REBELS on brighter and more massive systems. The fit to our simulated galaxies is consistent with a non-evolving linear slope. Our results are consistent with recent studies on the evolution of the galaxy main sequence at $z < 4 - 6$ \citep{Popesso+2022, Daddi+2022}, in that \texttt{dustyGadget} simulations sample the low-mass end of the MS, below the time-dependent turn-over mass that defines the transition between efficient and relatively inefficient star formation;
    \item a similarly good agreement is found when comparing the redshift evolution of the specific star formation rate predicted by the simulation with a recent analysis that includes REBELS sources out to $z \sim 7$;
    \item the relation between stellar and dark matter halo mass predicted by the simulations shows a large scatter, particularly at the low-mass end, likely reflecting the large variety of galaxy assembly histories. We find a broad agreement with some observational determinations \citep{Finkelstein+2015, Stefanon+2021}, and with models that assume a redshift independent relation \citep{Tacchella+2018}, indicating a constant star formation efficiency for a given halo mass across the redshift range we have investigated. However, at $z \le 6$ the simulated galaxies appear to have stellar masses that grow more efficiently than predicted by abundance matching methods \citep{Behroozi_2019, SunFurlanetto+2016}, particularly at the low-mass end.  This resonates with the large number density of galaxies at the faint-end of the stellar mass function, at Log$(\rm{M_\star/M_\odot}) < 9$. Assessing the statistical relevance of these galaxies with JWST observations will be fundamental to characterise their properties and correctly model these environments in future simulations (Venditti et al., in prep.);
    \item dust and stellar mass are confirmed to be related with the s-shape relation found in \citet{Graziani+2020} which changes its derivative for objects with stellar masses Log($\rm{M_{\star}/M_{\odot}}) \geq 8.5$. According to our simulations, dust enrichment at $z > 4$ is driven by stellar dust production and ISM grain growth, with the latter mechanism providing a growing contribution at the high-mass end. Overall, we find a good agreement with dust mass determinations for ALPINE galaxies by \citet{Pozzi+2021} at $4 \le z \le 6$ and with REBELS galaxies with stellar masses ${\rm Log(M_\star/M_\odot)} \leq 9.5 $ at $6 < z \le 8$ \citep{Sommovigo+2022},  while more massive REBELS galaxies appear to have dust masses systematically lower than our predictions, indicating that either our simulated high-mass galaxies are too dusty or that the dust temperature for some REBELS sources may have been over-estimated. Similar conclusions apply when the comparison is made with a subsample of ALPINE galaxies recently analysed by \citet{Sommovigo2022alpine}. Interestingly, the recent detection of dust continuum from 3 galaxies at $6 < z \leq 7$ by \citet{Witstok+2022} provides additional indications on their dust-to-stellar mass relation, with two galaxies being consistent with \texttt{dustyGadget} predictions and one galaxy showing a very large dust-to-stellar mass ratio, implying a very efficient dust production mechanism in this system, beyond what is predicted by \texttt{dustyGadget} for galaxies with comparable stellar mass.
\end{itemize}

In summary, the stellar mass assembly and related scaling relations investigated in this manuscript indicate that on the cosmological scale \texttt{dustyGadget} prescriptions are in reasonable global agreement with current high-redshift data, including JWST ERO and ERS. 
The new set of simulations reveals, on the other hand, an interesting population of evolving galaxies with stellar masses in $8.0 \leq \rm Log(M_{\star}/ \rm M_{\odot}) \leq 9.0$. The number density of these galaxies exceeds some of the current observational estimates and model predictions for the stellar mass function at $z \le 6$, and are at the origin of the large scatter found in the halo mass-stellar mass relation. This indicates that galaxies hosted in DM halos with similar mass may experience different SFHs and chemical enrichment timescales, as also reflected in the dust-to-stellar mass relation, where galaxies with $\rm{Log(M_{\star}/ \rm M_{\odot}}) \sim 8.5$ at $z \le 6 - 7$ are characterized by a broad range of dust masses, with differences of up to 1.5 dex. We plan to explore some of these aspects in a forthcoming publication.

The impressive capabilities of the JWST already revealed by the early release observations will certainly shed some light on the relevance and physical properties of these low-mass objects, providing invaluable constraints to future theoretical models investigating the details of their ISM and their impact on cosmic reionization.  

\section*{Acknowledgments}
The authors would like to thank the anonymous referee for her/his useful suggestions. We thank S. Tacchella, S. Finkelstein, R. Bouwens, M. Topping, P. Oesch and G. Popping for useful discussions. LG and RS acknowledge support from the Amaldi Research Center funded by the MIUR program "Dipartimento di Eccellenza" (CUP:B81I18001170001). 
We have benefited from the public available programming language \texttt{Python}, including the \texttt{numpy}, \texttt{matplotlib} and \texttt{scipy} packages.

\section*{Data Availability}
The data underlying this article will be shared on reasonable request to the corresponding author.
\bibliographystyle{mn2e}
\bibliography{dustyEnvsArticle}

\begin{appendix}
\section{SMF comparison with other simulations}
\label{sec:AppSMF}

\begin{figure*}
\centering
\includegraphics[width=\textwidth]{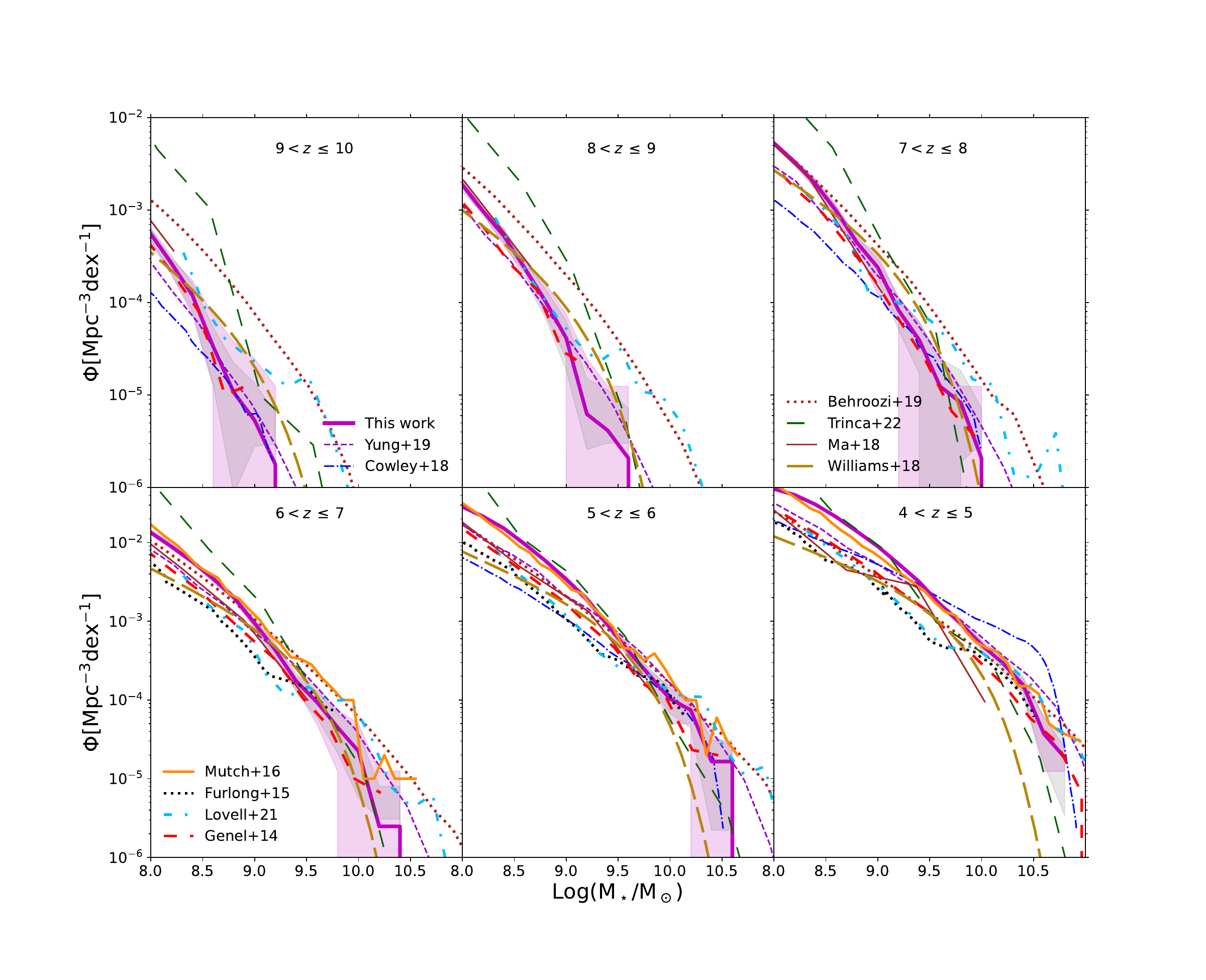}
 \caption{Mass Function obtained from our simulated galaxies in the range $z=10-4$: \textit{magenta solid lines} show mean values of our sample, \textit{pink shaded areas} the min-max spread of the simulations and \textit{gray shaded areas} the Poissonian error associated with our sample. The results of the semi-analytic model by \citet{Yung+2019} are shown as \textit{purple dashed lines} and those of the \texttt{CAT} \citep{Trinca+2022} semi-analytic model are in \textit{green loosely dashed}, the results of the \texttt{UNIVERSEMACHINE} model by \citet{Behroozi_2019} are in \textit{dark-red dotted lines} and the predictions by \citet{Cowley+18} in \textit{dash-dotted blue lines}. \textit{Olive dashed lines} are the predictions from the \texttt{JAGUAR} model by \citet{Williams+2018}, \textit{red dashed lines} are from the work by \citet{Genel+2014} (\texttt{Illustris}) and the \textit{dashed-dotted lightblue lines} are the predictions from the \texttt{FLARES} model \citep{Lovell+2021,Wilkins+2022}. Finally, \textit{solid brown lines} are predictions from the \texttt{FIRE-2} simulation \citealt{Ma+2018}, \textit{black dotted lines} are from the \texttt{EAGLE} simulation by \citet{Furlong+2015} and \textit{orange solid lines} are from the \texttt{DRAGONS} simulation by \citet{Mutch+2016}.}

\label{fig:FigSMFmod}
\end{figure*}

In this Section we compare \texttt{dustyGadget} results with independent model predictions\footnote{Some of the relevant data has been taken from the public repository at \url{https://github.com/stephenmwilkins/flags_data}}. The results are illustrated in Figure \ref{fig:FigSMFmod}, where we have applied the proper conversion factors \citep{Madau&Dickinson_2014} to stellar masses that were originally computed with an IMF different from the \citet{Salpeter+1955} one.
 
Semi-analytic forecasts from \citet{Yung+2019}\footnote{\url{https://www.simonsfoundation.org/semi-analytic-forecasts-for-jwst/}} are shown as purple dashed lines, while similar estimates based on the \texttt{GALFORM} code combined with a large DM simulation provided by  \citet{Cowley+18} are shown as blue dash-dotted lines. Dotted red lines and loosely dashed green lines are respectively the predictions from the \texttt{UNIVERSEMACHINE}\footnote{The \texttt{UNIVERSEMACHINE} (\url{https://www.peterbehroozi.com}) applies simple empirical models of galaxy formation to dark matter halo merger trees. It keeps track of two stellar masses: the "true" M$_{\star}$ given by the integral of past star formation minus stellar mass loss, and the "observed" M$_{\star}$ which includes systematic offsets and scatter as a function of redshift. Here we consider the "observed" stellar mass.} \citep{Behroozi_2019} and from the \texttt{CAT} semi-analytic model by \citet{Trinca+2022}. Dashed olive lines are from the phenomenological model \texttt{JAGUAR} \citep{Williams+2018}. This model is based on observed stellar mass and UV luminosity functions that have been measured in the redshift range $0 < z < 10$. The red dashed lines are from the N-body/hydrodynamical \texttt{Illustris-1} simulation by \citet{Genel+2014}, brown solid and dash-dotted lightblue lines are from cosmological zoom-in simulations, respectively from \texttt{FIRE-2} (version 2.0 of the \texttt{FIRE} project; \citealt{Hopkins+2018}) by \citet{Ma+2018} and \texttt{FLARES} \citep{Lovell+2021,Wilkins+2022}. Predictions from the \texttt{EAGLE} hydrodynamical simulations \citep{Furlong+2015} are in black dotted lines, while those from the \texttt{DRAGONS} semi-analytic galaxy formation model \citep{Mutch+2016} are in orange solid lines.

The comparison among the SMFs predicted by different models shows a large scatter, that is mostly due to different star formation conditions/feedback implementations in each model. The scatter is particularly evident at high-redshifts (top three panels) while it decreases at $z < 6 - 7$ (bottom three panels). At $7 < z \le 10$ \texttt{dustyGadget} simulations are in better agreement with the predictions by \citet{Yung+2019} and \citet{Genel+2014}, while they foresee a higher (lower) number of objects compared to \citet{Cowley+18} (\texttt{UNIVERSEMACHINE}, \citealt{Behroozi_2019}, and \texttt{CAT}, \citealt{Trinca+2022}). At $z \le 7$, our predictions at the low-mass end are in excellent agreement with the results of \citet{Mutch+2016}, and converge to the number densities predicted by \texttt{CAT} at $z \le 5$, but exceed the other model predictions. At all redshifts, the high-mass end ($\rm{Log(M_\star/M_\odot)} > 10 $) of the relation shows a large scatter between the models, and our simulations can not constrain the SMF for $\rm{Log(M_\star/M_\odot)} > 10$, due to the limited number of galaxies predicted within the simulated volumes in this mass range.

Figure \ref{fig:FigSMF} shows that current observations start to constrain the SMF at $z < 7$, and future JWST data will provide invaluable indications on the physics of star formation and feedback to be implemented in galaxy evolution models.

\section{MS comparison with other simulations}
\label{sec:AppMS}
\begin{figure*}
\centering
\includegraphics[width=\textwidth]{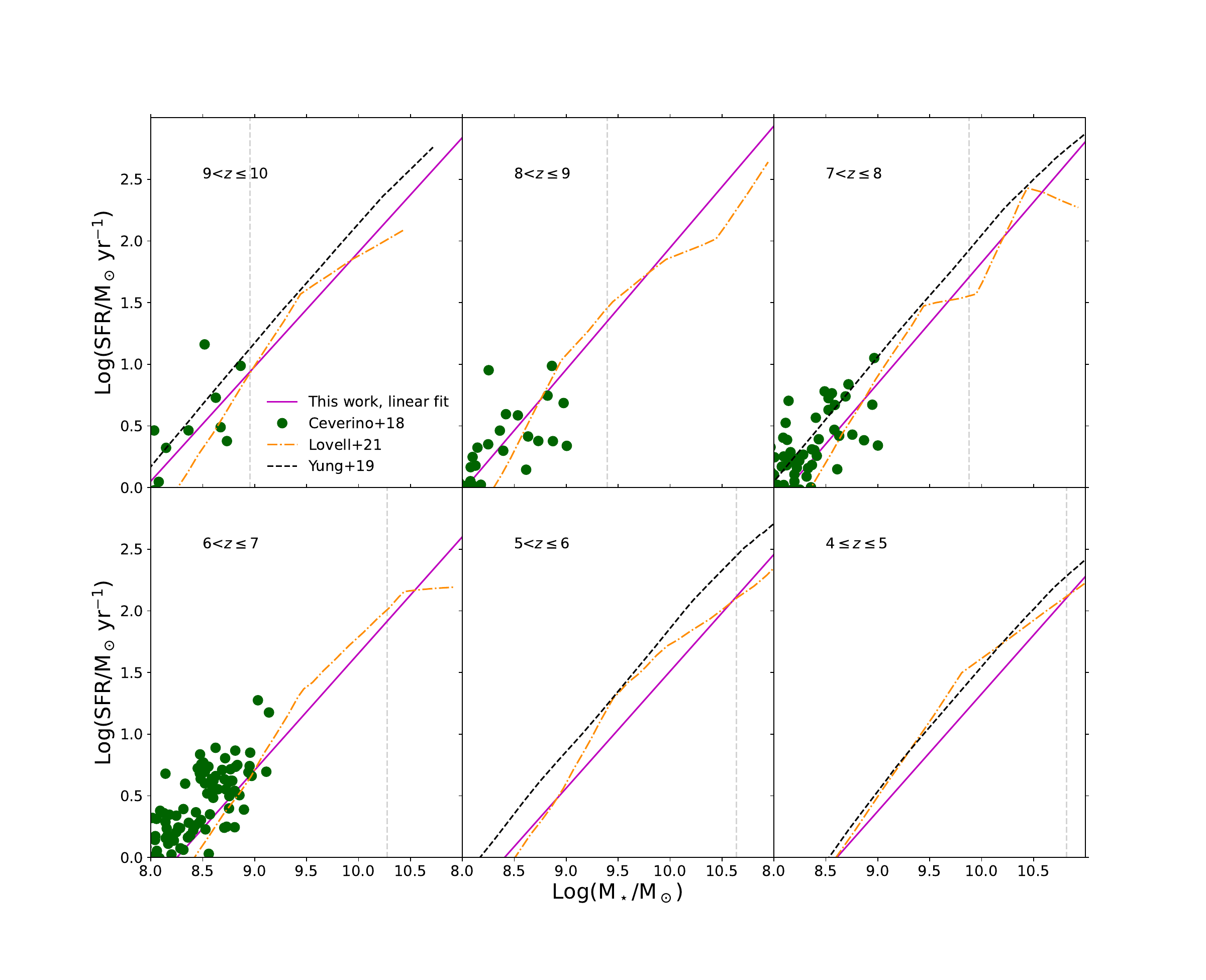}
\caption{Comparison between the MS predictions from our work and other simulations. \textit{Magenta lines} are our MS linear fit, \textit{green dots} show galaxies from the \texttt{FirstLight} project \citep{Ceverino+2017,Ceverino+2018}, \textit{dash-dotted orange lines} show the predictions from the \texttt{FLARES} simulation \citep{Lovell+2021} and \textit{dashed black lines} are from the semi-analytic model of \citet{Yung+2019}.}
\label{fig:FigMS_sim}
\end{figure*}

To strengthen the reliability of our results, \texttt{dustyGadget} results are also compared with predictions from  independent semi-analytical and numerical simulations. Wherever necessary, we use the conversion factors by \citet{Madau&Dickinson_2014} to convert results based on different IMF assumptions to the Salpeter IMF \citep{Salpeter+1955} adopted in our simulations. In particular, we compared our predictions with the zoom-in simulations by the \texttt{FLARES} \citep{Lovell+2021} and  \texttt{FirstLight} \citep{Ceverino+2017,Ceverino+2018} projects. We also compare our results with the semi-analytic predictions by \citet{Yung+2019}, based on a slightly modified version of the Santa Cruz model to sample halos over a wide mass range. \citet{Yung+2019} adopt a merger tree algorithm based on the Extended Press-Schechter formalism and, at each  redshift, they set up a grid of root halos spanning a certain range in virial velocity and assign them their expected volume-averaged abundances. Finally, for each root halo in the grid, they generated one-hundred Monte Carlo realizations of the merger histories. Comparing the predictions of different simulations is certainly not a straightforward task because of the different strategies each simulation adopts. Vertical dashed lines in Figure \ref{fig:FigMS_sim} show the $\rm{M_{\star,max}}$ found in our RefRun (U6), meaning that the fit above this mass has to be interpret as an extrapolation. Despite their different physical assumptions and simulation techniques, a good agreement is found among model predictions at $\rm{Log(M_\star/M_\odot) \leq Log(M_{\star,max}/M_\odot)}$, as already discussed by \citet{Graziani+2020}. At larger masses, FLARES simulations find a piecewise fit to the star forming sequence \citep{Lovell+2021}, and \citet{Yung+2019} find a change in slope at the high-mass end, to account for the bending seen in the sequence \citep{Popesso+2022, Sandles+2022}. Our simulations do not sample these high-masses, and our extrapolated fit does not predict a change of slope (see also the discussion in Section \ref{sec:MSslopeevo}).

\end{appendix}

\label{lastpage}
\end{document}